\documentclass[lettersize,journal]{IEEEtran}
\ifCLASSOPTIONcompsoc
\usepackage[nocompress]{cite}
\else
\usepackage{cite}
\fi

\usepackage{algorithmic}
\usepackage{array}
\usepackage{textcomp}
\usepackage{stfloats}
\usepackage{xcolor}
\definecolor{wV}{HTML}{08306b}
\definecolor{wV-DNN}{HTML}{323251}
\definecolor{woV}{HTML}{FFFFE0}
\usepackage{tikz}
\usepackage{pgfplots}
\pgfplotsset{compat=1.18}
\usetikzlibrary{patterns}
\usepackage{verbatim}
\usepackage{algorithm,amsbsy,amsmath,amsfonts,amssymb,epsfig,bbm,mathrsfs,multirow,amsthm,xcolor,cite}
\usepackage{epstopdf}
\usepackage[hidelinks]{hyperref}
\usepackage{graphicx,caption,subcaption}
\usepackage{setspace}
\usepackage[labelformat=simple]{subcaption}

\newcommand{\argmax}{\mathop{\rm argmax}}

\PassOptionsToPackage{hyphens}{url}
\usepackage{url}

\begin{document}

\title{Collaborative Learning Framework to Detect Attacks in Transactions and Smart Contracts}

\author{Tran Viet Khoa, Do Hai Son, Chi-Hieu Nguyen, Dinh Thai Hoang, Diep N. Nguyen, Tran Thi Thuy Quynh, \\ Trong-Minh Hoang, Nguyen Viet Ha, Eryk Dutkiewicz, Mohammad Abu Alsheikh, and Nguyen Linh Trung. \vspace*{-1cm}

\thanks{T.~V.~Khoa and M.~A.~Alsheikh are with the University of Canberra, Australia (e-mail: \{khoa.tran, mohammad.abualsheikh\}@canberra.edu.au).} 

\thanks{C.~H.~Nguyen, D.~T.~Hoang, D.~N.~Nguyen, and E.~Dutkiewicz are with the School of Electrical and Data Engineering, University of Technology Sydney, Sydney, NSW 2007, Australia (e-mail: hieu.c.nguyen@student.uts.edu.au, \{hoang.dinh, diep.nguyen, eryk.dutkiewicz\}@uts.edu.au).}
\thanks{D.~H.~Son is with the VNU Information Technology Institute, Hanoi, Vietnam (e-mail: dohaison1998@vnu.edu.vn).}
\thanks{N.~L.~Trung, T.~T.~T.~Quynh, and N.~V.~Ha are with the Advanced Institute of Engineering and Technology (AVITECH), University of Engineering and Technology, Vietnam National University, Hanoi, Vietnam (e-mail: \{linhtrung, quynhttt, hanv\}@vnu.edu.vn).} 

\thanks{Trong-Minh Hoang is with the Posts and Telecommunications Institute of Technology, Vietnam (e-mail: hoangtrongminh@ptit.edu.vn).} 
}

\maketitle
\begin{abstract}

With the escalating prevalence of malicious activities exploiting vulnerabilities in blockchain systems, there is an urgent requirement for robust attack detection mechanisms. To address this challenge, this paper presents a novel collaborative learning framework designed to detect attacks in blockchain transactions and smart contracts by analyzing transaction features. Our framework exhibits the capability to classify various types of blockchain attacks, including intricate attacks at the machine code level (e.g., injecting malicious codes to withdraw coins from users unlawfully), which typically necessitate significant time and security expertise to detect. To achieve that, the proposed framework incorporates a unique tool that transforms transaction features into visual representations, facilitating efficient analysis and classification of low-level machine codes. Furthermore, we propose an advanced collaborative learning model to enable real-time detection of diverse attack types at distributed mining nodes. Our model can efficiently detect attacks in smart contracts and transactions for blockchain systems without the need to gather all data from mining nodes into a centralized server. In order to evaluate the performance of our proposed framework, we deploy a pilot system based on a private Ethereum network and conduct multiple attack scenarios to generate a novel dataset. To the best of our knowledge, our dataset is the most comprehensive and diverse collection of transactions and smart contracts synthesized in a laboratory for cyberattack detection in blockchain systems. Our framework achieves a detection accuracy of approximately 94\% through extensive simulations and 91\% in real-time experiments with a throughput of over 2,150 transactions per second. 
These compelling results validate the efficacy of our framework and showcase its adaptability in addressing real-world cyberattack scenarios.

\end{abstract}



\section{Introduction}\label{sec:Int}
\IEEEPARstart{B}{lockchain} technology has been rapidly being developed with many applications in recent years. This technology was initially developed with a well-known digital currency application named Bitcoin. After that, many potential applications using this technology have been developed beyond cryptocurrency. The tremendous development of this technology is from the fact that it provides a new approach to data sharing and storage without the need for any third party (e.g., bank and government). Blockchain is a decentralized environment in which transactions and smart contracts can be recorded and executed in a secure and transparent manner. It is challenging to manipulate transactions once they are put into the blocks. Thus, blockchain technology protects data integrity, and its applications have been widely developed in various fields of industry, such as smart manufacturing, supply chain management, the food industry, smart grid, healthcare, and the Internet of Things~\cite{huo2022comprehensive, zuo2023survey, Kang2024, Cheng2023}.

Smart Contracts (SCs) are solely programs in blockchain systems (e.g., Ethereum and Solana). 
SCs define and enforce a set of rules for users via using codes. They also facilitate users' interactions by allowing them to send transactions to execute defined functions. 
In practical scenarios, attackers can inject malicious codes into SCs and transactions before deployment to attack a blockchain system for specific purposes. By default, SCs and their interactions are irreversible after deployment in a blockchain system~\cite{buterin2014ethereum}. If the attacked SCs and transactions are validated in the blockchain network, the consequences are inevitable~\cite{wang2019blockchain}. For instance, SCs exhibit various vulnerabilities~\cite{swc}, which attackers can exploit to engage in injurious purposes, including unauthorized coin withdrawals from other users' pockets and taking control of the system~\cite{hassan2022anomaly, wang2019blockchain, Leng2022, jiang2022exploring}. Specifically, in 2016, a smart contract (SC) named Decentralized Autonomous Organization (DAO) was a victim of a re-entrancy attack. At that time, this SC held \$150~million in the Ethereum network, and this attack led to a hardfork of Ethereum that created Ethereum Classic~(ETC)~\cite{swc}. In addition, the 4Chan group created an SC named Proof of Weak Hands Coin (PoWHC) on the Ethereum system. However, this SC witnessed an underflow attack that caused a loss of 866~ETH (i.e., Ethereum coins)~\cite{powh}. Although most of the attacks in blockchain systems happened in the finance sector, many blockchain-based applications have been developing in different sectors such as healthcare, supply chain, and food industry~\cite{Yuan2018, Li2019}. Therefore, securing blockchain systems against cyber threats has become an imperative necessity. 

There are a number of challenges to detect and prevent attacks in transactions and SCs. The first challenge is the lack of a dataset synthesized in the laboratory for various kinds of attacks on transactions and SCs in a blockchain system. In recent research (e.g.,~\cite{huang2021hunting} and~\cite{chen2021defectchecker}), the authors use datasets from the public blockchain network and label data using the attack records history. When using this method to label attack data, it is assumed that the benign data does not include the attacks. Therefore, generating data, which has ``clean'' samples (i.e., transactions between users without any malicious behavior) of normal behavior and attacks in transactions and SCs, is urgently needed. However, a blockchain system in the mainnet has large and diverse types of data. Thus, a synthesized dataset from the laboratory needs to be diverse and similar to reality.
The second challenge is to understand and analyze the content of Bytecode, the compiled form of an SC's source code. It is worth noting that the main functions of the transactions and SCs are encoded into the Bytecode, which is represented by a series of hexadecimal numbers, to be implemented in a blockchain system~\cite{wang2019blockchain}.
It is crucial for a real-time attack detection system to analyze the content of Bytecode to detect attacks in a blockchain system~\cite{huang2021hunting}. There are two approaches to analyze the Bytecode, i.e., using the source code of SCs for comparison and analyzing the Bytecode.
Unfortunately, only 1\% source codes of SCs are open~\cite{huang2021hunting}, and analyzing Bytecode without the corresponding source code of SCs and transactions can be unreliable and time-consuming~\cite{huang2021hunting}.
The third challenge is that most of the current attack detection models are centralized. Thus, they need to gather all data (i.e., transactions together with their labels, e.g., attack or normal) into a centralized model to perform training and testing. However, blockchain systems are decentralized environments so it is challenging to collect data from all mining nodes (MNs) to perform training at the centralized server. In addition, if we transfer data from all MNs to the centralized server for processing (e.g., training and testing), data privacy can be compromised.

Given the above, in this paper, we first set up experiments in our laboratory to deploy various kinds of attacks on transactions and SCs in a blockchain system (i.e., a private Ethereum system). These attacks were recorded as occurring in the real world and resulted in serious consequences for the Ethereum system. To address the first challenge, we collect all the transactions in MNs to build a novel dataset, called~\textbf{Blockchain Transaction-based Attacks Dataset (BTAT)}\footnote{https://avitech-vnu.github.io/BTAT}. To the best of our knowledge, this is the first cyberattack dataset on transactions and SCs in a blockchain network synthesized in a laboratory. To enrich the dataset, we create a large number of individual accounts to send transactions to the blockchain network for execution randomly.
This dataset can be used for both research and industry purposes to address cyberattacks in transactions and SCs. In addition, to deal with the second challenge of Bytecode analysis, we propose a novel ML-based framework that analyzes transactions and SCs without the need to understand the SC source codes. The main goal of our proposed framework is to detect attacks in SCs after their deployment in a blockchain network, and before such transactions are validated and added to the main chain. 
Our proposed framework automatically extracts transaction features in real-time and efficiently analyzes them to detect attacks.
To achieve real-time analysis, we collect unvalidated transactions (i.e., pending transactions) in a blockchain system and analyze them to detect attacks before they are validated and added to the main chain.
To facilitate this, we first build a highly effective tool, called~\textbf{Blockchain Code Extraction and Conversion Tool (BCEC)}, to convert important information of pending transactions and SCs to an image form. This tool calls the transaction using a transaction hash (i.e., a feature of the transaction) and then extracts key fields like Bytecode and value from the transactions. After that, it can convert the contents into images for further processing. 
Second, we develop an ML-based approach based on CNN to learn and detect attacks for transactions and SCs. To the best of our knowledge, \textbf{this is the first ML-based framework that analyzes the Bytecode directly and detects various types of attacks in transactions and SCs}. Such an ML-based framework, which uses important information from transactions for analysis, is more flexible and easier to detect new types of attacks than other vector-based methods. 
To address the third challenge about centralized attack detection, we develop a highly-effective collaborative cyberattack detection framework that can detect cyberattacks inside transactions and SCs in real-time with high accuracy. In our proposed framework, the CNN of each mining node can exchange learning knowledge (i.e., the trained models) with other nodes to create a global model. In this way, the learning model of each node can improve the detection accuracy without sending their local data over the network. Our major contributions can be summarized as follows:

\begin{itemize} 

	\item We implement a blockchain system and perform experiments to build a novel dataset named BTAT. To the best of our knowledge, this is the first dataset with cyberattacks on transactions and SCs of a blockchain system that is synthesized in a laboratory. 

    \item We develop BCEC that can collect transactions, extract their features, and convert them into images to build a dataset. This tool can be implemented in real-time to support the analysis of the attack detection framework.

	\item We develop a real-time attack detection framework that can be deployed at the mining nodes to detect attacks in transactions and SCs for a blockchain network. In our framework, the mining nodes can detect attacks in transactions and SCs in real-time at about 2,150 transactions per second.
 
    \item We propose a collaborative learning framework that can efficiently detect attacks in blockchain networks. In our framework, each mining node can exchange learning knowledge with others and then aggregate a new global model without any centralized model. In this way, our framework can achieve high accuracy in detecting attacks without exposing the mining node's local dataset over the network.    
 
	\item We perform both simulations and real-time testing to evaluate our proposed framework. Our proposed framework can achieve accuracy up to 94\% in simulation and 91\% in real-time experimental results. In addition, our framework has the capacity to analyze various types of transaction features, expanding the detection capabilities for the diversity of attacks.   
\end{itemize} 

\section{Related work}\label{sec:Relatedwork}
There are several works trying to deal with attacks on transactions and SCs in blockchain networks.
In~\cite{qian2020towards}, the authors propose to convert the source codes of SCs into vectors. They then use bidirectional long-short-term memory to identify abnormal patterns of vectors to detect re-entrancy attacks. The simulation results show that their proposed model can achieve 88.26\% F1-Score and 88.47\% accuracy in detecting re-entrancy attacks. 
In~\cite{wang2020contractward}, the authors propose to detect the vulnerabilities inside SCs. To do this, they use feature extraction to analyze the Bytecode of SCs. 
In this paper, the authors use various types of machine learning models to detect 6 types of vulnerabilities with an F1-score of up to 97\%. Even though the methods in~\cite{qian2020towards, wang2020contractward} can detect some types of attacks, they need to use source code of SCs in high-level programming languages (e.g., Solidity). It is worth noting that when an SC is created, the SC creates corresponding transactions for execution and then sends them to MNs for the mining process. From the MN point of view, we only can observe transactions with the encoded content (e.g., Bytecode) in their features. In real-time attack detection, we need to analyze this content to find out the attacks in transactions and SCs.

\begin{figure*}[t!]
    \hspace{2.9cm}
	\includegraphics[width=0.85\linewidth]{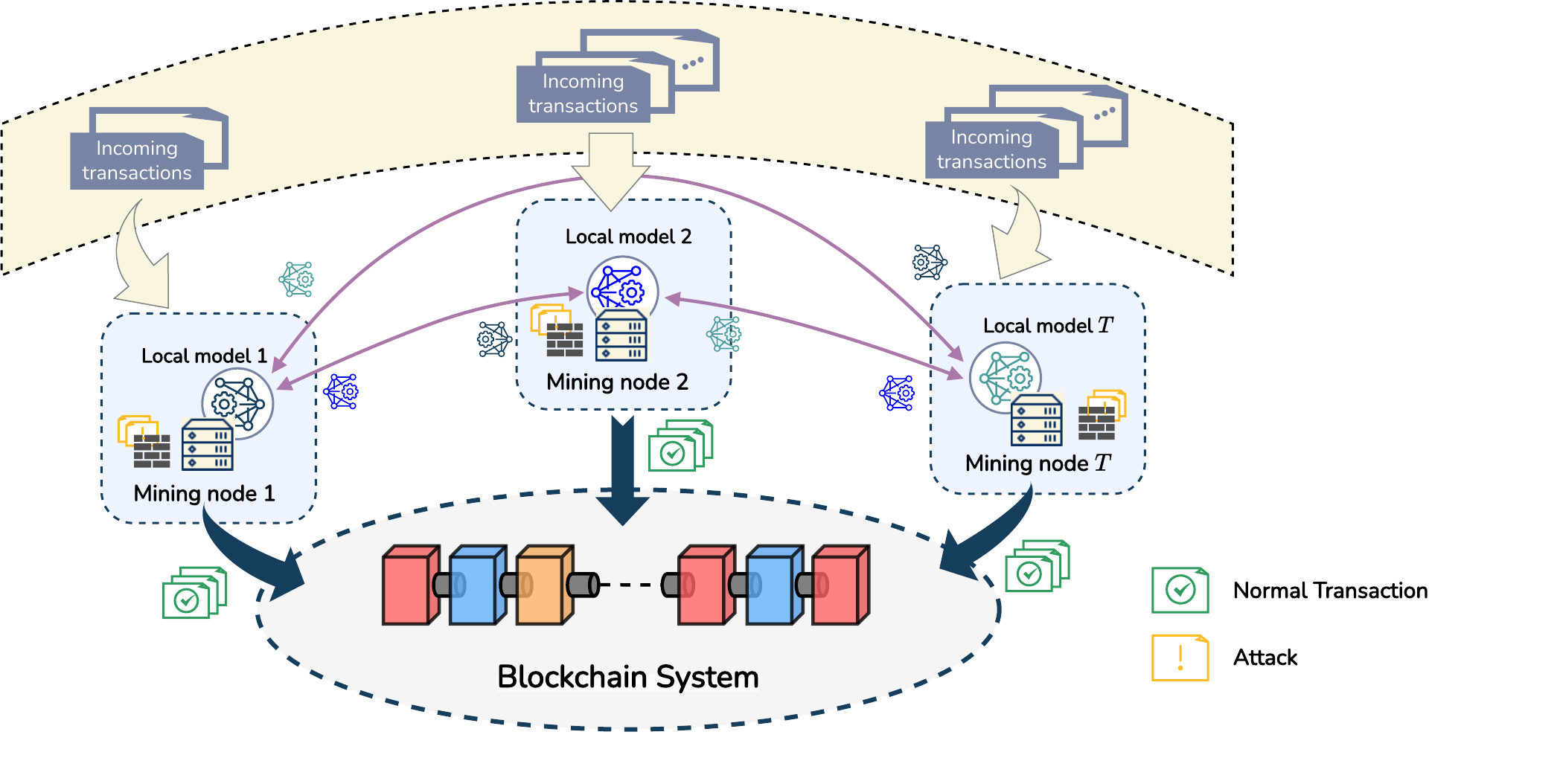}
	\caption{The system model of our proposed framework. While receiving transactions, our framework will perform preprocessing to extract important information. After that, our collaborative learning will perform the attack detection process to detect network normal behavior or a type of attack.}
	\label{fig:System}
    \vspace*{-0.3cm}
\end{figure*}

Unlike the above deep learning approaches, 
in~\cite{wang2019contractguard}, the authors propose an intrusion detection system named ContractGuard that can defend Ethereum SCs against malicious attacks. The experimental results show that ContractGuard successfully guards against attacks on all real-world vulnerabilities without the need of Solidity source code. However, ContractGuard requires additions of 36.14\% gas consumption and 28.27\% time consumption compared to those of the original SCs. Moreover, another trade-off of the proposed scheme is that it needs centralized administrations to manually verify anomaly transactions.
In~\cite{nguyen2019detect}, the authors study the Bytecode. They propose to use the attack vector method to directly analyze the Bytecode. 
This approach can effectively detect some specific attacks by using pre-defined vectors.
However, this method is difficult to extend to various types of new attacks. In addition, even though the attack detection ability can achieve up to 100\% in some types of attacks (e.g., re-entrancy, delegatecall, overflow, etc), the authors only test this method in a small scale of data (about 100 samples). 
In~\cite{huang2021hunting}, the authors propose to use Graph embedding to analyze Bytecode. To do this, the authors convert the Bytecode of SC into vectors and then compare the similarities between the vectors of SC to detect the attacks of SC. The experimental results show that this method can achieve a precision of up to 91.95\% in detecting attacks. 
Both~\cite{nguyen2019detect} and~\cite{huang2021hunting} have to use source code to analyze the bytecode. In~\cite{Ivanov2023}, the authors introduce an SC security testing approach with the aim of identifying the suspicious behaviors associated with vulnerabilities of SCs in blockchain networks. According to their evaluations, the proposed framework completely rejected about 3.5\% of transactions due to being untestable. Therefore, they point out that further Bytecode analysis can reduce this portion.
In addition, in~\cite{chen2021defectchecker}, the authors propose DefectChecker to analyze vulnerabilities in SCs. This framework uses symbolic execution to analyze Bytecode without the need for source codes. This framework can detect eight types of vulnerabilities in SCs and get an F1-score of 88\%. 

All of the methods above focus on centralized learning. To implement those methods, all the data needs to be gathered in a centralized server for learning and analysis. However, blockchain is a decentralized environment and MNs are distributed worldwide. Thus, gathering all blockchain data to perform training and testing is impractical. In this paper, we propose a collaborative learning model framework that can detect attacks for SCs and transactions without the need of a centralized server in blockchain networks. Differing from the above works, we introduce an innovative ML-based framework to analyze Bytecode directly to detect attacks inside transactions without the need for source code. To do this, we propose to convert the encoded information of transactions into images.
Our proposed framework can analyze these images to detect various types of attacks in both transactions and SCs. In this way, our proposed framework is flexible and can effectively detect new types of attacks. 

\section{Blockchain System: Fundamental and Proposed Collaborative Learning framework}\label{sec:Sys}

\subsection{Blockchain}

Blockchain technology is a decentralized method to store and manage data. In a blockchain system, each MN can be used to store and process data. When a Mining Node (MN) receives transactions, it typically groups them into a block as a part of the mining process. However, it is worth noting that the consensus mechanism is responsible for managing the rules of the mining process in a blockchain network. There are various types of consensus mechanisms being used in blockchain networks~\cite{Lashkari2021}. For example, Ethereum 2.0 uses Proof-of-Stake (PoS)~\cite{Buterin2020} as its consensus mechanism for the mining process. In PoS, a validator, who is responsible for proposing a new block, is randomly selected based on the amount of staked ETH in users' deposits. When the mining process is completed, the valid block is added to the main chain of blocks. After that, the block is irreversible to ensure the integrity of transactions in a blockchain. Another characteristic of blockchain is transparency which enables all MNs to access the history of transactions within a blockchain network. This transparency ensures total transaction records are visible to all MNs and promotes trust in the blockchain network. Overall, blockchain possesses numerous valuable characteristics, including decentralization, transparency, immutability, and data tamper resistance, making it applicable across various sectors to enhance human life.

\subsection{Designed Blockchain System and Our Proposed Collaborative Learning Framework}

In our laboratory, we set up experiments to collect datasets for training and testing our framework. We first deploy a blockchain system based on a private Ethereum network in our laboratory (more details are shown later in Section~\ref{sec:setting}). This network uses the latest version of the Ethereum network (i.e., Ethereum 2.0). This version uses PoS as a consensus mechanism for validating new blocks. Our system includes various MNs to collect data from their local networks and bootnodes, the management nodes to connect MNs together. The MNs can receive transactions from various types of blockchain applications such as smart cities, smart agriculture, IoT, and cryptocurrency. As described above, the transactions are first sent to MNs. They are then put into a block, and the MNs will perform the mining process to put them into the main chain. We perform various attacks using malicious transactions and SCs on this system. These attacks (i.e., DoS with block gas limit, overflows and underflows, flooding of transactions, re-entrancy, delegatecall, and function default visibility) happened and caused serious damage to blockchain systems~\cite{Chen2020}. Through experiments, we build a state-of-the-art dataset with both normal and attacked transactions and SCs to evaluate the performance of attack detection methods.

In this paper, we consider a blockchain system with $T$ MNs working in a blockchain system as described in Fig.~\ref{fig:System}. When an MN receives pending transactions from the blockchain network, it uses~\textbf{BCEC} to preprocess them by extracting information from important features and then converting them to grey images. After that, we propose a collaborative learning framework for analyzing the images to detect attacks in transactions and SCs. In our framework, each MN uses its local dataset to train a deep neural network. After the training process, each MN shares its trained model with other nodes and also receives their trained models in return. Afterward, every MN aggregates all the received trained models from other nodes together with its current trained model to generate a new global model for further training (we will explain more details in the next section). In this way, MN can exchange its learning knowledge with the neural network of other MNs. This approach can not only improve the overall learning knowledge of the neural network of all MNs but also protect the privacy of local data over network transmission. By preventing the transmission of the local data of each MN over the network, our approach can also reduce network traffic to avoid network congestion. Thus, the neural networks of MNs can improve the accuracy of detecting attacks for transactions and SCs in blockchain systems.

\section{Proposed Attack Detection framework}

\begin{figure*}[t!]
	\centering
	\includegraphics[width=\linewidth]{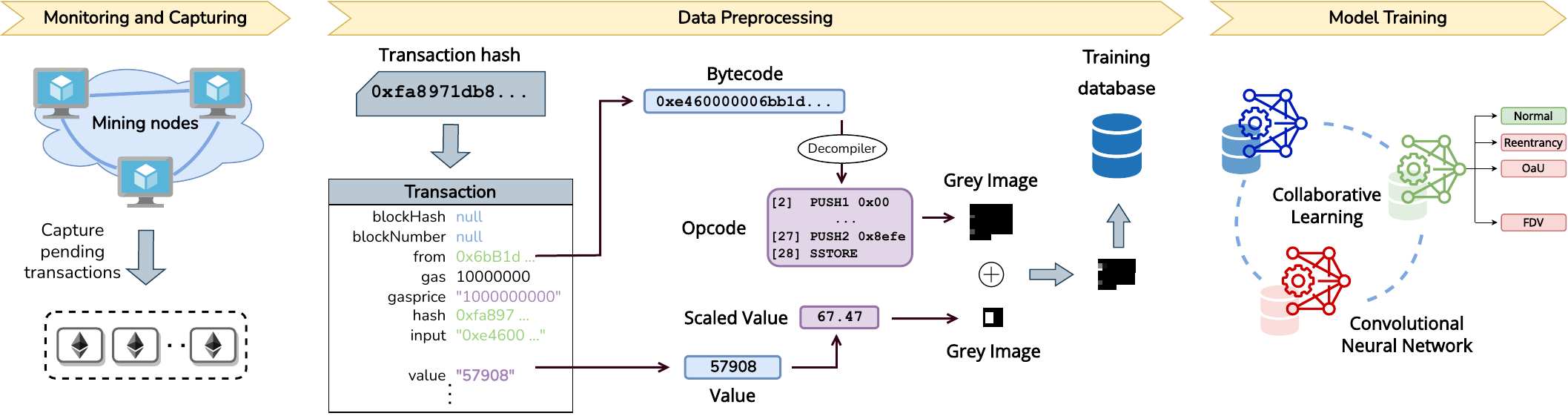}
	\caption{The processes of our proposed framework: The captured pending transactions are preprocessed using the BCEC tool. This tool extracts important features of transactions and converts them into images. After that, the images are processed by the CNN and collaborative learning models to detect attacks.}
	\label{fig:pre}
    \vspace*{-0.5cm}
\end{figure*}

In our proposed attack detection framework, the MNs are used to learn and share their learning knowledge with others to improve the accuracy of their attack detection. At each MN, we propose to use a deep neural network as a detector to learn the data of the MN's local system. After that, the MN exchanges its learning knowledge (i.e., trained model) with other MNs. When an MN receives trained models from others, it will integrate these models with its current model to train its local dataset. This process is iteratively repeated until reaching a predefined number of iterations. In summary, our proposed framework includes three processes as described in Fig.~\ref{fig:pre}. The first process is preprocessing. In this process, our proposed framework captures and extracts the important information of the incoming transactions and then converts them to grey images. The second process is to develop a deep convolution neural network to classify the grey images to detect attacks. The last process is collaborative learning. In this process, each MN can exchange the trained model with others to improve the accuracy of attack detection. 

\subsection{Preprocessing Process}

Fig.~\ref{fig:pre} describes our proposed preprocessing process for transactions in a blockchain system. The main purposes of the preprocessing process are extracting the important features from incoming transactions and converting them into images for further processing. It is worth noting that SCs are a set of agreements to deploy transactions. For implementation, a server has to send transactions of the SCs to the MN for a mining process. From the MN point of view, we can only observe transaction hashes (i.e., the unique addresses of incoming transactions), which are represented in a series of hexadecimal numbers. The preprocessing process has three steps to deal with these transaction hashes as follows:

\begin{itemize}
\item \textbf{Step 1:} Capture transaction hashes from the MN and then recover pending transactions from transaction hashes to have the full information of all transaction features such as content, value, block hash, block number, chainID, etc.

\item \textbf{Step 2:} Extract the content of two crucial features in transactions named Bytecode and value. The bytecode feature includes the main functions of transactions and the value feature indicates the amount of ETH (Ethereum's native tokens) involved in a transaction.
Although we can effectively use the bytecode feature in detecting various types of attacks in transactions and SCs, 
it does not provide any information on some specific types of attacks, such as Flooding of Transactions~\cite{SCMA}, where the transaction content is null. Thus, it may be inefficient if we only rely on the bytecode feature for analysis. Therefore, we propose to enhance the attack detection framework by incorporating information from the value feature (its benefits are discussed in more detail in Section~\ref{sec:setting}). After that, we apply appropriate preprocessing methods to the corresponding features as follows:
\begin{itemize}
    \item \textbf{Bytecode feature}: Extract the content and then transform it into opcode using EVM Bytecode Decompiler~\cite{EVM_Bytecode}. The opcode is a series of executed comments in assembly. Thus, we propose to convert all features of this assembly code to a grey image named Grey Image 1.
    
    \item \textbf{Value feature}: we first scale its content to an appropriate range and then convert it to another grey image named Grey Image 2.
\end{itemize}

\item \textbf{Step 3:} In this step, we combine both Grey Image 1 and Grey Image 2 to create the Final Grey Image. This Final Grey Image includes all essential information of a transaction and an SC in the blockchain system. They can be used to train the deep convolution neural network to find out the attacks inside. 

\end{itemize}

In this framework, all these steps are encapsulated in the~\textbf{BCEC} tool. This tool can perform the preprocessing process in real-time to support the analysis of collaborative attack detection to detect attacks for transactions and SCs in a blockchain system.

\subsection{Learning Process}

In our proposed framework, at each MN, we implement a detector that can help to detect attacks based on the transformed images from the preprocessing process with high accuracy. The core component of the detector is developed based on a Deep Convolutional Neural Network (CNN). The reason for using CNN is that this framework can classify a large amount of labeled data, especially in image classification with high accuracy~\cite{rawat2017deep}. Additionally, in our proposed approach, the CNN model does not have to learn their local data separately, it can exchange its trained model with other MNs to improve the learning knowledge as well as enhance the accuracy of attack detection. In detail, the architecture of CNN in an MN includes three types of layers, i.e., convolution layer, max pooling layer, and fully connected layer~\cite{rawat2017deep}. These layers are described as follows:


\begin{itemize} 
	\item \textbf{Convolution layer:} The neurons in this layer are formed in feature maps to learn the feature representation of the input. In addition, these feature maps can connect with others of the previous layer by weight parameters called filter banks~\cite{lecun2015deep}. In this layer, the input data is convoluted with weight parameters in every iteration to create feature maps.
	
	\item \textbf{Max pooling layer:} The main purpose of this layer is to reduce the resolution of feature maps in the previous layer. To do this, this layer selects the largest values in areas of feature map~\cite{rawat2017deep} and then sends them to the next layer.

	\item \textbf{Fully connected layer:} This layer performs classification functions for the neural network. In this layer, the feature maps from previous layers are first flattened. They are then put into a fully connected layer for classification. The softmax function is included at the end of this layer to produce the output prediction.
\end{itemize}

We denote $\boldsymbol{D}$ as a local dataset of an MN to train a CNN. $\boldsymbol{D}$ includes $\boldsymbol{S}$ images and $\boldsymbol{Y}$ labels so we can denote $\boldsymbol{D}=(\boldsymbol{S},\boldsymbol{Y})$. We consider $n=\{1,.., N\}$ as the training layer of the neural network. We denote $N$ as the number of training layers of the neural network. We denote $\boldsymbol{I}$ as the matrix features of image $\boldsymbol{S}$, and $\boldsymbol{I_{i}}$ as the matrix features of image $\boldsymbol{S}$ at iteration $i$. The output of a convolution layer $n$, $n \in \{1,.., N\}$, at iteration $i+1$ can be calculated as follows~\cite{saputra2022federated}:

\begin{equation}
\begin{aligned}
\label{eqn1}
\boldsymbol{I}_{n+1,i} = \gamma_n \Big(\boldsymbol{I}_{n,i} * \boldsymbol{F}_n\Big),
\end{aligned}
\end{equation}
where $(*)$ is the convolutional operation, $\gamma_n$ is the activation function and $\boldsymbol{F}_n$ is the filter bank of layer $n$. After that, the output of the convolution layer is put into a max pooling layer. The output of a max pooling layer can be calculated as follows:
\begin{equation}
\begin{aligned}
\label{eqn2}
\boldsymbol{I}_{n+2,i} = \varphi \Big(\boldsymbol{I}_{n+1,i}\Big),
\end{aligned}
\end{equation}
where $\varphi$ is the max pooling function that selects the maximum value in a pooling area. We denote $\boldsymbol{I}_{e,i}$ as the matrix features of the last image after processing with multiple convolution layers and max pooling layers. $\boldsymbol{I}_{e,i}$ is put into a softmax function to classify and produce the output in the fully connected layer. We consider $l\in \{1,...,L\}$ as the classification group number, $\hat{Y}_l \in \boldsymbol{\hat{Y}}$ as the output prediction, the probability that an output prediction $\hat{Y}$ belongs to group $l$ can be calculated as follows:
\begin{equation}
\begin{aligned}
\label{eqn3}
p(\hat{Y}_l = l|\boldsymbol{I}_{e,i},\boldsymbol{W}_{e,i},\boldsymbol{b}_{e,i}) &= softmax(\boldsymbol{W}_{e,i},\boldsymbol{b}_{e,i}) \\
&= \frac{\exp{(\boldsymbol{W}_{e,i}\boldsymbol{I}_{e,i} + \boldsymbol{b}_{e,i})}}{\sum_l \exp{(\boldsymbol{W}_{e,l,i}\boldsymbol{I}_{e,i} + \boldsymbol{b}_{e,l,i})}},
\end{aligned}
\end{equation}
where $\boldsymbol{W}_{e,i},\boldsymbol{b}_{e,i}$ are the weights and biases of the fully connected layer at iteration $i$, respectively; and $\boldsymbol{W}_{e,l,i},\boldsymbol{b}_{e,l,i}$ as weights and biases of the fully connected layer at iteration $i$ to classify an output prediction into class $l$. Based on equation~(\ref{eqn3}), we can calculate a vector of prediction $\boldsymbol{\hat{Y}}$ which includes output prediction $\hat{Y}_l$ belonging group $l$ with probability $p$ as follows:
\begin{equation}
\begin{aligned}
\label{eqn4}
\boldsymbol{\hat{Y}}= \argmax_l [p(\hat{Y}_l = l|\boldsymbol{I}_{e,i},\boldsymbol{W}_{e,i},\boldsymbol{b}_{e,i})].
\end{aligned}
\end{equation}

In this stage, we compare the output predictions with the labels using a sparse categorical cross-entropy function to calculate the loss for backpropagation. We denote $Y_l \in \boldsymbol{Y}$ as the label of class $l$ in $\boldsymbol{Y}$. The loss function can be calculated as follows:
\begin{equation}
\begin{aligned}
\label{eqn5}
\boldsymbol{J}(\boldsymbol{W})= -\sum_{l=1}^L Y_l \log \hat{Y}_l.
\end{aligned}
\end{equation}

We denote $\boldsymbol{W}$ as the model of the neural network. Based on equation~(\ref{eqn5}), we can calculate the gradient of this function as follows:
\begin{equation}
\begin{aligned}
\label{eqn6}
\nabla \boldsymbol{\theta} &= \frac {\partial \boldsymbol{J}(\boldsymbol{W})}{\partial \boldsymbol{W}} 
&= -\frac {\partial \Big(\sum_{l=1}^L Y_l  \log \hat{Y}_l \Big)}{\partial \boldsymbol{W}}.
\end{aligned}
\end{equation}

After having the gradient based on equation~(\ref{eqn6}). We then use it for the Adam optimizer to update the parameters of the neural networks. We consider $m_{i+1}$ and $v_{i+1}$ as the moment vectors of the next iteration $i+1$ of the Adam optimizer. The $m_{i+1}$ and $v_{i+1}$ can be calculated from the gradient and Adam functions~\cite{kingma2014adam} as $m_{i+1}= A_1 (\nabla \boldsymbol{\theta})$ and $v_{i+1}= A_2 (\nabla \boldsymbol{\theta})$. We denote $\boldsymbol{\Gamma}_{i}$ as a trained model, and ${\theta}_i$ as a global model at iteration $i$. With $\beta_{i+1}$ as the learning rate, a new trained model at the next iteration $i+1$ can be calculated as follows:
\begin{equation}
\begin{aligned}
\label{eqn7}
\boldsymbol{\Gamma}_{i+1} &= \boldsymbol{\Gamma}_i - \beta_{i+1} \frac {m_{i+1}} {\sqrt{v_{i+1}}} \\
&= \boldsymbol{\Gamma}_i - \beta_{i+1} \frac {A_1 (\nabla \boldsymbol{\theta}_i)} {\sqrt{A_2 (\nabla \boldsymbol{\theta}_i)}}.
\end{aligned}
\end{equation}
\subsection{Collaborative Learning Process}

In this paper, we propose a Collaborative Deep Convolutional Neural Network framework (Co-CNN) to detect the different types of attacks in a blockchain network. In this framework, each MN has a CNN model to train and test its dataset. The CNN model can receive trained models from other MNs to improve the accuracy of attack detection. To do this, the CNN model of an MN first gets the trained model (gradient) based on equation~(\ref{eqn6}). It then sends the trained model to other MNs and receives trained models from others. We denote $T$ as the total number of MNs and $t \in T$ as the MN number. We consider at iteration $i$, an MN receives $T-1$ trained models from others. $\boldsymbol{\theta}_{t,i}$ is the trained model of MN $t$ at iteration $i$. It can aggregate all trained models using the following formula~\cite{Federated_learning1}:
\begin{equation}
\begin{aligned}
\label{eqn8}
\boldsymbol{\theta}_{i+1} = \frac{1}{T}\sum_{t=1}^T\boldsymbol{\theta}_{t,i},
\end{aligned}
\end{equation}
where $\boldsymbol{\theta}_{i+1}$ is the new aggregated trained model. After generating a new aggregated trained model, each MN will calculate a new trained model using equation~(\ref{eqn7}). This process continuously repeats until the algorithm converges or reaches the predefined maximum number of iterations. After the training process, we can obtain the optimal trained model in each MN to analyze and detect the attacks inside a series of grey images. This process is summarized in Algorithm~\ref{al:Classification_FL}.

\begin{algorithm}
	\algsetup{linenosize=\tiny}
	\caption{The learning process of Co-CNN model}
	\label{al:Classification_FL}
	\begin{algorithmic}[1]
	    \WHILE{$i$ $\leq$ maximum number of iterations}
	        \FOR{$\forall t \in T$}
        		\STATE The CNN of the MN-$t$ learns $D_t$ to produce $\hat{Y}$. 
        		\STATE The MN-$t$ creates gradient $\theta_t$ and sends it to others
        		\STATE The MN-$t$ receives $T-1$ gradients from others.		        
                \STATE MN calculates a new optimal trained model $\boldsymbol{\Gamma}_{i+1}$. 
            \ENDFOR	
            \STATE {$i=i+1$.}
		\ENDWHILE
		\STATE MN uses its optimal model $\boldsymbol{\Gamma}_{optimal}$ to detect attacks based on input grey images.
	\end{algorithmic}
\end{algorithm}

\section{Experiment and Performance Analysis}\label{sec:setting}

\subsection{Experiment Setup}\label{sec:exper_setup}

In our experiments, we set up an Ethereum 2.0 system in our laboratory, as shown in Fig.~\ref{fig:exp_model}. This version of Ethereum uses a new consensus mechanism, namely PoS instead of Proof-of-Work~(PoW).
In our experiments, there are five Ethereum nodes, two bootnodes, a trustful device, and an attack device. All these devices are connected to a Cisco switch, which serves as the central hub for our local network. The configuration of these devices is as follows:

\begin{figure}
    \centering
    \includegraphics[width=\linewidth]{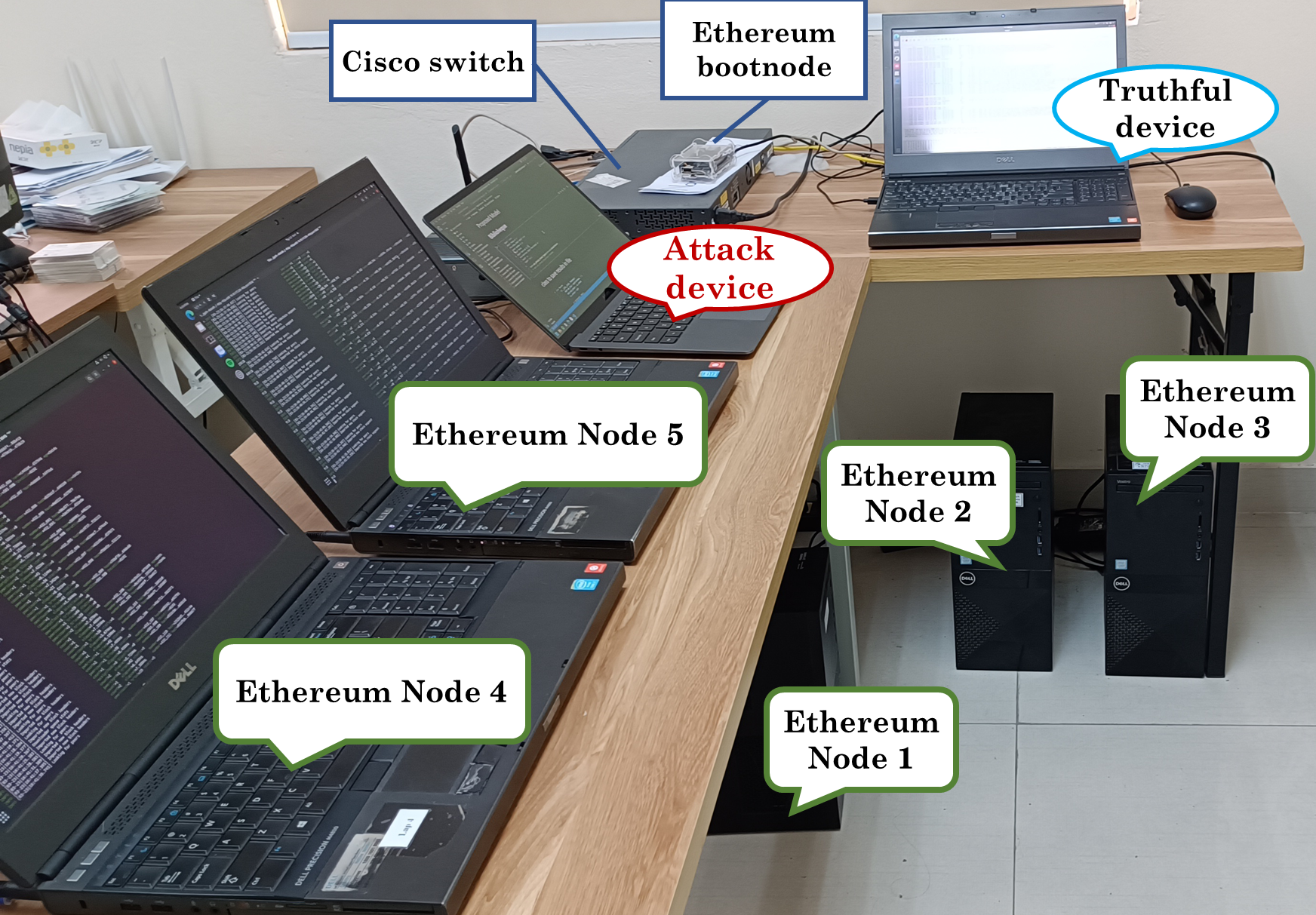}
    \caption{Real-time experiment setup.}
    \label{fig:exp_model}
    \vspace*{-0.5cm}
\end{figure}

\begin{itemize}
    \item Ethereum nodes are launched by \textit{Geth~v1.10.22} - an official open-source implementation of Ethereum network~\cite{Go-Ethereum} and \textit{Prysm~v3.2.0} - an official implementation of the PoS consensus mechanism in Ethereum~2.0~\cite{Prysm}. They share the same genesis configurations, e.g., chainID, block gas limit at 30,000,000 gas, etc. The configurations of nodes 1, 2, and 3 are workstation computers with processor Intel~Core~i9-10900~@5.2~GHz, RAM of 64~GB. The configurations of nodes 4 and 5 are personal computers with processor Intel~Core~i7-4810MQ~@3.8~GHz, RAM of 16~GB.
        
    \item \textit{Geth} bootnode and \textit{Prysm} bootnode are also created by \textit{Geth~v1.10.22} and \textit{Prysm~v3.2.0}, respectively. They are responsible for connecting all the Ethereum nodes together.
\end{itemize}

\subsection{Dataset Collection}
According to the detailed analysis of the public Ethereum network on transaction behavior~\cite{Anwar2021}, the addresses that are associated with less than $10$ transactions account for $88$\% of total addresses. About $50$\% received addresses appear only one time for a transaction in history. This is because most people want to create transactions anonymously. Therefore, to create diversity and reality for our dataset, we need to create a large number of unique accounts (i.e., 10,000 accounts in our experiments) to send transactions to Ethereum nodes. A truthful server, as shown in Fig.~\ref{fig:exp_model}, randomly selects accounts from these accounts to create transactions for the blockchain system.

\subsubsection{Normal State}
For the normal state, we use \textit{OpenZeppelin Contracts}~\cite{zeppelin} library as the secured SCs. Two types of transactions below are used to generate samples randomly for the normal state.

\begin{itemize}
    \item Exchange ETH: On the public Ethereum network, most transactions only exchange the ETH to another address without any bytecode. This kind of transaction accounts for $75$\% of the total samples of the normal state in our experiment.
    
    \item Transactions-related SCs: There are two types of these transactions. The transactions for deploying SCs and the transactions that interact with functions in deployed SCs.
    We perform three essential SCs' categories in the Ethereum system, i.e., Tokens/Coins/NFT, Ethereum 2.0 deposit, and SCs for other purposes.
    
\end{itemize}

Although the number of original SCs is minuscule compared to the total transactions in the dataset. The content of transactions and deployed SCs are not duplicated. The reason is that we randomly select not only the senders and recipients but also the amount of ETH and inputs of functions in any generated transaction. 

\subsubsection{Attack States}

SCs have a number of vulnerabilities listed in SWC~\cite{swc} because of programmers, consensus mechanisms, and compilers. Attackers can exploit these weaknesses of SC to perform attacks and then steal money in blockchain systems~\cite{Chen2020}. In this work, we regenerate several real-world attacks from the tracks that they left on Ethereum's ledger. We give a brief description of the six types of application layer-based attacks.
\begin{table}
\centering
\caption{Number of samples on the proposed BTAT dataset.}
\label{tab:dataset}
    \begin{tabular}{|l|c|c|} 
    \hline
    \multicolumn{1}{|c|}{\textbf{Class}} & \textbf{Number of samples} & \textbf{Portion (\%)} \\
    \hline
    Normal & 152,423 & 50.34 \\ 
    \hline
    DoS & \phantom{0}22,994 & \phantom{0}7.59 \\ 
    \hline
    OaU & \phantom{0}29,254 & \phantom{0}9.66 \\ 
    \hline
    FoT & \phantom{0}41,732 & 13.78 \\ 
    \hline
    Re & \phantom{0}22,682 & \phantom{0}7.49 \\ 
    \hline
    DeC & \phantom{0}22,455 & \phantom{0}7.41 \\ 
    \hline
    FDV & \phantom{0}11,209 & \phantom{0}3.73 \\ 
    \hline
    \textit{Total} & 302,749 & \phantom{0.}100 \\
    \hline
    \end{tabular}
\end{table}
\begin{itemize}
    \item \textit{DoS with Block Gas Limit (DoS)}: There are several functions inside SCs. These functions can be temporarily disabled when their gas requirements exceed the block gas limit. A \textit{DoS} case occurred in 2015 when SC GovernMental's 1,100 ETH jackpot payout was stuck~\cite{swc}. The GovernMental SC is deployed in our work, and we continuously join the jackpot to disable the payout function. 

    \item \textit{Overflows and Underflows (OaU)}: In solidity language,
    if a variable is out of its range, it is in the overflow or underflow state. In this case, the variable is turned to another value (e.g., $0$ for overflow and $2^{256}-1$ for underflow). 
    Attackers can use this vulnerability to bypass SCs' conditions when withdrawing funds. For example, they can bypass the requirements of checking their accounts' balances. Several real \textit{OaU} attacks were detected, e.g., $2^{256}$ BEC tokens, CSTR token, USD~\$$800$k of PoWH token~\cite{powh}, and so on~\cite{swc}. We re-perform the above \textit{OaU} attacks on their original SCs in the dataset. 

    \item \textit{Flooding of Transactions (FoT)}: Attackers spam a number of meaningless transactions to delay the consensus of blockchain networks. Such an attack caused the unconfirmation of $115$k Bitcoin transactions in 2017~\cite{SCMA}. In our setup, \textit{FoT} attacks are generated by continuously sending a negligible amount of ETH from a random sender to another arbitrary recipient.
    
    \item \textit{Re-entrancy (Re)}: When the SCs do not update their states before sending funds, attackers can recursively call the withdraw function to drain the SCs’ balances. Two types of \textit{Re} are single-function and cross-function. The single-function type happened and led to a loss of 3.6 million ETH in 2016. Both types of Re are performed in our dataset~\cite{swc}.
    
    \item \textit{Delegatecall (DeC)}: \textit{delegatecall()} is the mechanism to inherit functions, storage, and variables from other deployed SCs.
    If the inherited SCs are attacked, they will in-directly affect the main SC. To implement, we re-create the $2^{\text{nd}}$ Parity MultiSig Wallet attack~\cite{swc}. In this attack, attackers took control and suicide the inherited SC.
    
    \item \textit{Function Default Visibility (FDV)}: 
    If the programmers do not define the visibility of functions in SCs, it will default to the public. Thus, anyone can interact with those functions. For implementation, we perform the $1^{\text{st}}$ Parity MultiSig Wallet attack~\cite{swc}. In this attack, attackers took control of this SC through an \textit{FDV} flaw.
\end{itemize}

\begin{figure*}
    \centering
    \begin{subfigure}[b]{0.36\textwidth}
         \centering
         \includegraphics[width=\textwidth]{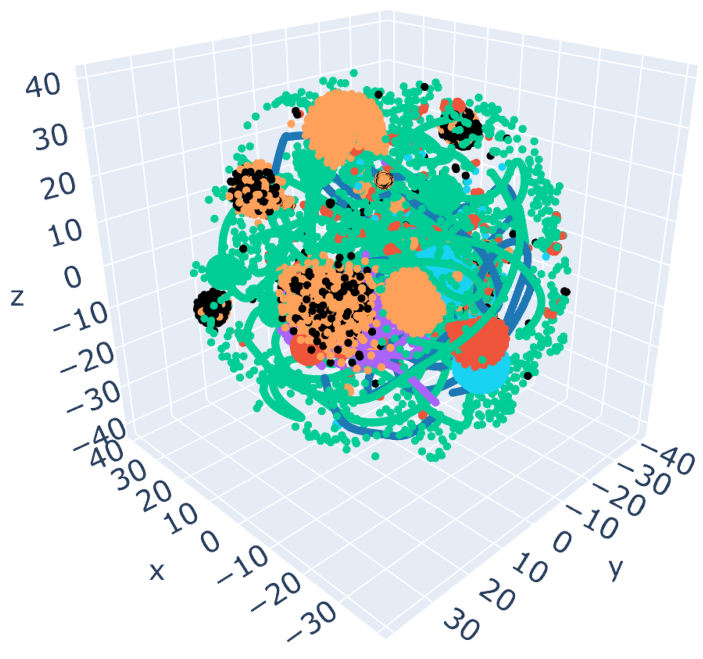}
         \caption{}
         \label{fig:3d}
     \end{subfigure}
     \hfill
     \begin{subfigure}[b]{0.44\textwidth}
         \centering
         \includegraphics[width=\textwidth]{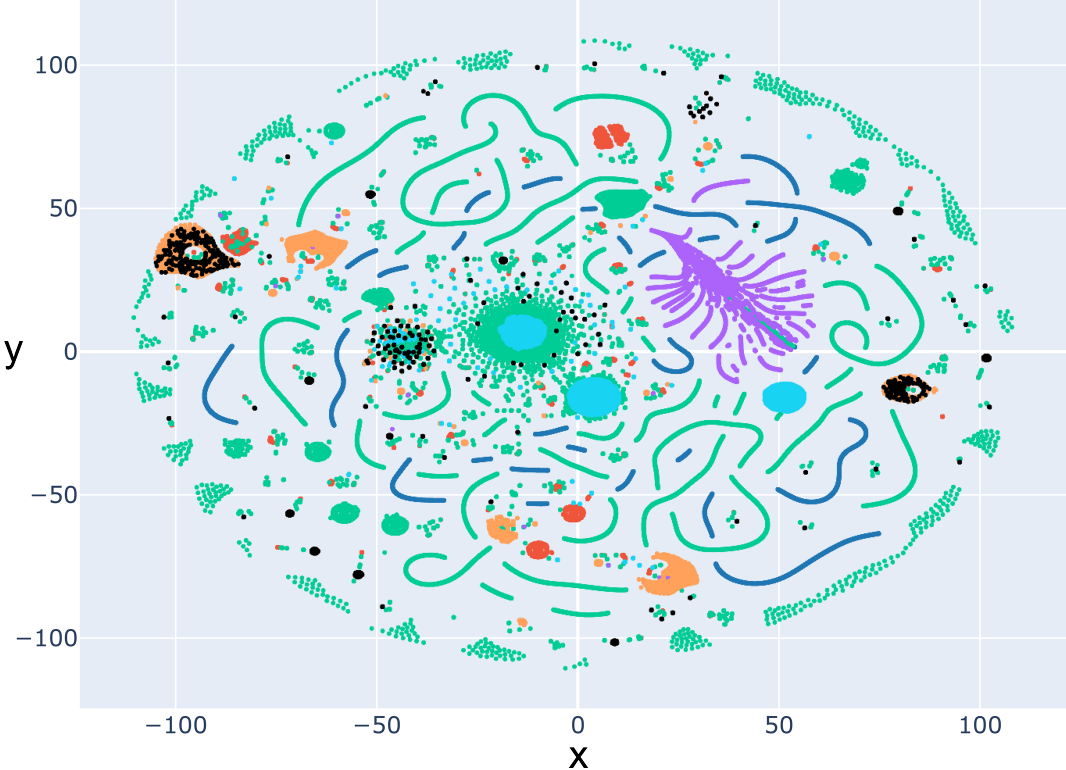}
         \caption{}
         \label{fig:2d}
    \end{subfigure}
    \hfill
    \begin{subfigure}[b]{0.5\textwidth}
         \centering
         \includegraphics[width=\textwidth]{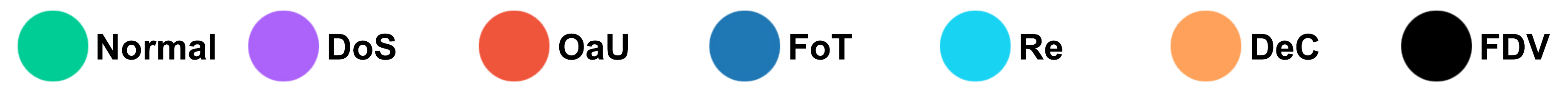}
    \end{subfigure}
    \hfill
    \caption{Visualization using t-SNE of BTAT dataset: (a) Visualization in 3D. (b) Visualization in 2D.}
    \label{fig:tsne}
    \vspace*{-0.3cm}
\end{figure*}

Table~\ref{tab:dataset} shows the number of samples in each class of our proposed dataset. The proportions of the samples in the classes are not balanced, e.g., the number of \textit{Re} samples is twice that of \textit{FDV}. Because \textit{Re} requires a series of attack transactions instead of only one attack transaction as in \textit{FDV}. We use~\textit{t}-Distributed Stochastic Neighbor Embedding ~(\textit{t}-SNE)~\cite{tsne} to visualize our designed BTAT in 3D and 2D, as shown in Fig.~\ref{fig:tsne}. In this figure, the points of all classes are randomly scattered and form non-linear lines. Additionally, we can see in Fig.~\ref{fig:2d} that numerous points of~\textit{FDV} and~\textit{DeC} overlap with each other, and the same scenario occurs with~\textit{Normal} and~\textit{Re}. These overlaps create significant challenges for attack detection in the next section.

\subsection{Evaluation Methods}
\begin{figure}
    \centering
    \resizebox{.49\textwidth}{!}{%
    \begin{tikzpicture}
      \begin{axis}[
        xlabel={},
        ylabel={Percentage},
        set layers,
        ybar=5pt,
        x=3cm,
        bar width=12pt,
        symbolic x coords={Accuracy, Precision, Recall},
        grid,
        yminorgrids=true,
        ymin=0, ymax=100,
        enlarge x limits={0.3},
        xtick = {Accuracy, Precision, Recall},
        legend cell align=left,
        nodes near coords,
        nodes near coords style = {font=\fontsize{6}{6}\selectfont},
        ylabel near ticks, ylabel shift={-5pt},
        legend style={at={(80pt, -50pt)},anchor=south,draw=none, style={column sep=0.5cm}},
        minor tick num=3,
        legend columns=2,
        ylabel style={font=\bfseries},
        xtick pos=left, xtick style={draw=none}
        ]
        \addplot[black,fill=woV,postaction={pattern=north east lines}] coordinates {
          (Accuracy, 72.16) (Precision, 58.91) (Recall, 58.64)
        };
        \addlegendentry{Centralized-CNN w/o-V}
        \addplot[black,fill=red,postaction={pattern=grid}] coordinates {
          (Accuracy, 72.48) (Precision, 60.74) (Recall, 46.22)
        };
        \addlegendentry{Centralized-Decision Tree w/-V}
        \addplot[black,fill=wV-DNN,postaction={pattern=north west lines}, pattern color=white] coordinates {
          (Accuracy, 85.87) (Precision, 71.30) (Recall, 74.64)
        };
        \addlegendentry{Centralized-DNN w/-V}
        \addplot[black,fill=wV,postaction={pattern=dots}, pattern color=white] coordinates {
          (Accuracy, 93.85) (Precision, 90.41) (Recall, 89.74)
        };
        \addlegendentry{Centralized-CNN w/-V}
      \end{axis}
    \end{tikzpicture}
    }
    \caption{The results of the preprocessing processes in different schemes.}
    \label{fig:preprocessing}
    \vspace*{-0.5cm}
\end{figure}

The confusion matrix~\cite{confusion_matrix1, confusion_matrix2} is widely used to evaluate the performance of machine learning models. We denote TP, TN, FP, and TN as ``True Positive'', ``True Negative'', ``False Positive'', and ``True Negative''. In this paper, we use ubiquitous parameters (i.e., accuracy, precision, recall) in the confusion matrix to evaluate the performance of models. The accuracy of a model can be calculated as follows:
\begin{equation}
\begin{aligned}
\label{eqn9}
	\mbox{Accuracy} = \frac{\mbox{TP}+\mbox{TN}}{\mbox{TP}+\mbox{TN}+\mbox{FP}+\mbox{FN}}.
\end{aligned}
\end{equation}

In addition, we use the macro-average precision and macro-average recall to evaluate the performance of the models. With $L$ as the number of classification groups (i.e., the total number of normal and attack states), the macro-average precision is calculated as follows:
\begin{equation}
\begin{aligned}
\label{eqn10}
	\mbox{Precision} = \sum_{l=1}^L\frac{\mbox{TP}_l}{\mbox{TP}_l+\mbox{FP}_l}.
\end{aligned}
\end{equation}

The macro-average recall of the total system can be calculated as follows:
\begin{equation}
\begin{aligned}
\label{eqn11}
	\mbox{Recall} = \sum_{l=1}^L\frac{\mbox{TP}_l}{\mbox{TP}_l+\mbox{FN}_l}.
\end{aligned}
\end{equation}

\subsection{Simulation and Experimental Results}\label{subsec:result}

In this section, we present the simulation and real-time experimental results of our experiments. In particular, we use the confusion matrix to evaluate our proposed model's performance (in terms of accuracy, precision, and recall) compared to the centralized model. 

\subsubsection{Preprocessing Analysis} 

\begin{figure*}
    \begin{subfigure}{0.5\linewidth}
        \centering
        \includegraphics[width=.8\linewidth]{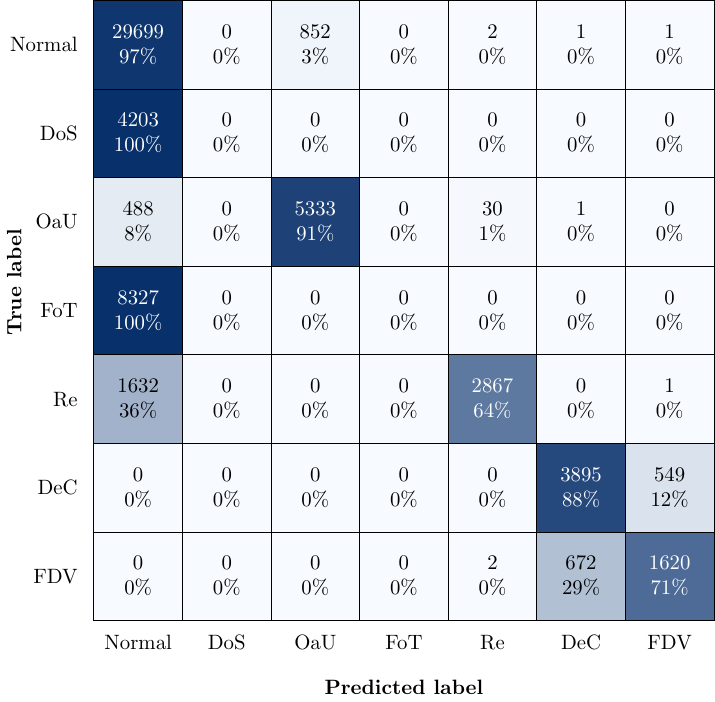}
        \caption{}
        \label{fig:wop}
    \end{subfigure}
    \begin{subfigure}{0.5\linewidth}
        \centering
        \includegraphics[width=.8\linewidth]{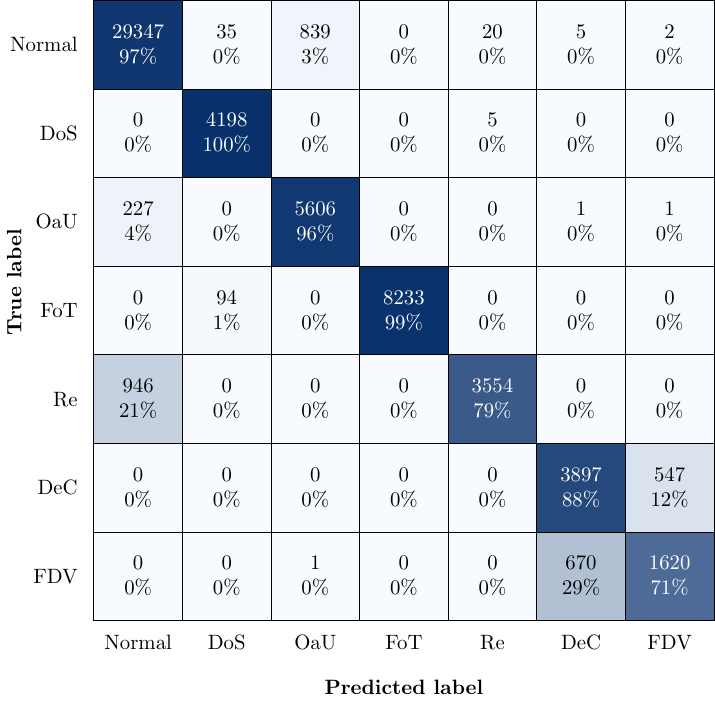}
        \caption{}
        \label{fig:wp}
    \end{subfigure}
    \caption{The detection results of the models w/ and w/o-V feature. (a) Centralized-CNN w/o-V. (b) Centralized-CNN w/-V.}
    \label{fig:wop_wp}
    \vspace*{-0.3cm}
\end{figure*}
\begin{table*}
\centering
\caption{Simulation results w/o-V with Centralized-CNN, Co-CNN-3, Co-CNN-5, and Co-CNN-10 models.}
\label{tab:simu_result_wop}
\begin{subtable}{\textwidth}
\centering
\resizebox{.7\textwidth}{!}{
\begin{tabular}{|l|c|ccc|ccccc|}
\hline
\multicolumn{1}{|c|}{\multirow{2}{*}{\textbf{}}} & \multirow{2}{*}{\textbf{Centralized-CNN}} & \multicolumn{3}{c|}{\textbf{Co-CNN-3}}                                                  & \multicolumn{5}{c|}{\textbf{Co-CNN-5}}                                                                                                                            \\ \cline{3-10} 
\multicolumn{1}{|c|}{}                           &                                       & \multicolumn{1}{c|}{\textbf{MN-1}} & \multicolumn{1}{c|}{\textbf{MN-2}} & \textbf{MN-3} & \multicolumn{1}{c|}{\textbf{MN-1}} & \multicolumn{1}{c|}{\textbf{MN-2}} & \multicolumn{1}{c|}{\textbf{MN-3}} & \multicolumn{1}{c|}{\textbf{MN-4}} & \textbf{MN-5} \\ \hline
\textbf{Accuracy}                                & 72.163                                & \multicolumn{1}{c|}{71.686}        & \multicolumn{1}{c|}{71.761}        & 72.080         & \multicolumn{1}{c|}{72.735}        & \multicolumn{1}{c|}{72.519}        & \multicolumn{1}{c|}{72.211}        & \multicolumn{1}{c|}{72.760}         & 72.627        \\ \hline
\textbf{Precision}                               & 58.911                                & \multicolumn{1}{c|}{58.323}        & \multicolumn{1}{c|}{58.298}        & 58.646        & \multicolumn{1}{c|}{59.676}        & \multicolumn{1}{c|}{59.300}          & \multicolumn{1}{c|}{58.818}        & \multicolumn{1}{c|}{59.699}        & 59.032        \\ \hline
\textbf{Recall}                                  & 58.638                                & \multicolumn{1}{c|}{57.539}        & \multicolumn{1}{c|}{57.951}        & 58.608        & \multicolumn{1}{c|}{58.955}        & \multicolumn{1}{c|}{58.807}        & \multicolumn{1}{c|}{58.415}        & \multicolumn{1}{c|}{59.444}        & 58.969        \\ \hline
\end{tabular}
}
\end{subtable}
\hfill
\vspace{0.2cm}

\begin{subtable}{\textwidth}
\centering
\resizebox{.7\textwidth}{!}{
\begin{tabular}{|l|cccccccccc|}
\hline
\multicolumn{1}{|c|}{\multirow{2}{*}{\textbf{}}} & \multicolumn{10}{c|}{\textbf{Co-CNN-10}}                                                                                                                                                                                                                                                                                                                    \\ \cline{2-11} 
\multicolumn{1}{|c|}{}                           & \multicolumn{1}{c|}{\textbf{MN-1}} & \multicolumn{1}{c|}{\textbf{MN-2}} & \multicolumn{1}{c|}{\textbf{MN-3}} & \multicolumn{1}{c|}{\textbf{MN-4}} & \multicolumn{1}{c|}{\textbf{MN-5}} & \multicolumn{1}{c|}{\textbf{MN-6}} & \multicolumn{1}{c|}{\textbf{MN-7}} & \multicolumn{1}{c|}{\textbf{MN-8}} & \multicolumn{1}{c|}{\textbf{MN-9}} & \textbf{MN-10} \\ \hline
\textbf{Accuracy}                                & \multicolumn{1}{c|}{72.768}        & \multicolumn{1}{c|}{73.333}        & \multicolumn{1}{c|}{73.184}        & \multicolumn{1}{c|}{73.117}        & \multicolumn{1}{c|}{73.150}         & \multicolumn{1}{c|}{72.984}        & \multicolumn{1}{c|}{73.017}        & \multicolumn{1}{c|}{73.267}        & \multicolumn{1}{c|}{73.516}        & 73.117         \\ \hline
\textbf{Precision}                               & \multicolumn{1}{c|}{58.169}        & \multicolumn{1}{c|}{59.462}        & \multicolumn{1}{c|}{59.107}        & \multicolumn{1}{c|}{58.957}        & \multicolumn{1}{c|}{58.779}        & \multicolumn{1}{c|}{58.621}        & \multicolumn{1}{c|}{58.288}        & \multicolumn{1}{c|}{59.503}        & \multicolumn{1}{c|}{59.013}        & 59.125         \\ \hline
\textbf{Recall}                                  & \multicolumn{1}{c|}{58.131}        & \multicolumn{1}{c|}{58.531}        & \multicolumn{1}{c|}{58.462}        & \multicolumn{1}{c|}{58.727}        & \multicolumn{1}{c|}{58.775}        & \multicolumn{1}{c|}{58.285}        & \multicolumn{1}{c|}{58.528}        & \multicolumn{1}{c|}{59.066}        & \multicolumn{1}{c|}{59.192}        & 58.650          \\ \hline
\end{tabular}
}
\end{subtable}
\end{table*}

\begin{table*}
\centering
\caption{Simulation results w/-V with Centralized-CNN, Co-CNN-3, Co-CNN-5, and Co-CNN-10 models.}
\label{tab:simu_result}
\begin{subtable}{\textwidth}
\centering
\resizebox{.7\textwidth}{!}{
\begin{tabular}{|l|c|ccc|ccccc|}
\hline
\multicolumn{1}{|c|}{\multirow{2}{*}{\textbf{}}} & \multirow{2}{*}{\textbf{Centralized-CNN}} & \multicolumn{3}{c|}{\textbf{Co-CNN-3}}                                                  & \multicolumn{5}{c|}{\textbf{Co-CNN-5}}                                                                                                                            \\ \cline{3-10} 
\multicolumn{1}{|c|}{}                           &                                       & \multicolumn{1}{c|}{\textbf{MN-1}} & \multicolumn{1}{c|}{\textbf{MN-2}} & \textbf{MN-3} & \multicolumn{1}{c|}{\textbf{MN-1}} & \multicolumn{1}{c|}{\textbf{MN-2}} & \multicolumn{1}{c|}{\textbf{MN-3}} & \multicolumn{1}{c|}{\textbf{MN-4}} & \textbf{MN-5} \\ \hline
\textbf{Accuracy}                                & 93.849                                & \multicolumn{1}{c|}{93.88}         & \multicolumn{1}{c|}{94.384}        & 94.115        & \multicolumn{1}{c|}{94.347}        & \multicolumn{1}{c|}{94.057}        & \multicolumn{1}{c|}{94.148}        & \multicolumn{1}{c|}{94.206}        & 94.439        \\ \hline
\textbf{Precision}                               & 90.413                                & \multicolumn{1}{c|}{90.216}        & \multicolumn{1}{c|}{91.162}        & 90.860         & \multicolumn{1}{c|}{90.794}        & \multicolumn{1}{c|}{90.540}         & \multicolumn{1}{c|}{90.637}        & \multicolumn{1}{c|}{90.903}        & 91.029        \\ \hline
\textbf{Recall}                                  & 89.742                                & \multicolumn{1}{c|}{89.665}        & \multicolumn{1}{c|}{90.688}        & 89.970         & \multicolumn{1}{c|}{90.329}        & \multicolumn{1}{c|}{89.932}        & \multicolumn{1}{c|}{90.025}        & \multicolumn{1}{c|}{90.514}        & 90.536        \\ \hline
\end{tabular}
}
\end{subtable}
\hfill
\vspace{0.2cm}

\begin{subtable}{\textwidth}
\centering
\resizebox{.7\textwidth}{!}{
\begin{tabular}{|l|cccccccccc|}
\hline
\multicolumn{1}{|c|}{\multirow{2}{*}{\textbf{}}} & \multicolumn{10}{c|}{\textbf{Co-CNN-10}}                                                                                                                                                                                                                                                                                                                    \\ \cline{2-11} 
\multicolumn{1}{|c|}{}                           & \multicolumn{1}{c|}{\textbf{MN-1}} & \multicolumn{1}{c|}{\textbf{MN-2}} & \multicolumn{1}{c|}{\textbf{MN-3}} & \multicolumn{1}{c|}{\textbf{MN-4}} & \multicolumn{1}{c|}{\textbf{MN-5}} & \multicolumn{1}{c|}{\textbf{MN-6}} & \multicolumn{1}{c|}{\textbf{MN-7}} & \multicolumn{1}{c|}{\textbf{MN-8}} & \multicolumn{1}{c|}{\textbf{MN-9}} & \textbf{MN-10} \\ \hline
\textbf{Accuracy}                                & \multicolumn{1}{c|}{93.633}        & \multicolumn{1}{c|}{94.248}        & \multicolumn{1}{c|}{93.849}        & \multicolumn{1}{c|}{93.566}        & \multicolumn{1}{c|}{93.899}        & \multicolumn{1}{c|}{93.832}        & \multicolumn{1}{c|}{93.516}        & \multicolumn{1}{c|}{93.732}        & \multicolumn{1}{c|}{93.699}        & 93.849         \\ \hline
\textbf{Precision}                               & \multicolumn{1}{c|}{89.326}        & \multicolumn{1}{c|}{90.611}        & \multicolumn{1}{c|}{90.095}        & \multicolumn{1}{c|}{89.969}        & \multicolumn{1}{c|}{90.106}        & \multicolumn{1}{c|}{90.048}        & \multicolumn{1}{c|}{89.252}        & \multicolumn{1}{c|}{90.684}        & \multicolumn{1}{c|}{89.778}        & 90.464         \\ \hline
\textbf{Recall}                                  & \multicolumn{1}{c|}{89.206}        & \multicolumn{1}{c|}{89.716}        & \multicolumn{1}{c|}{89.313}        & \multicolumn{1}{c|}{89.114}        & \multicolumn{1}{c|}{89.745}        & \multicolumn{1}{c|}{89.289}        & \multicolumn{1}{c|}{89.213}        & \multicolumn{1}{c|}{89.464}        & \multicolumn{1}{c|}{89.298}        & 89.477         \\ \hline
\end{tabular}
}
\end{subtable}
\end{table*}

In this section, we compare our proposed model in various schemes. On the one hand, we use our proposed preprocessing process as in Fig.~\ref{fig:pre} under different schemes such as CNN, Deep Neural Network (DNN), and Decision Tree (DT) to compare their evaluation results.  
On the other hand, we eliminate the value feature and use only the Bytecode preprocessing and the CNN to analyze the transactions and SCs. Through the results of these schemes, we demonstrate the efficiency of our proposed preprocessing process in combining various features of transactions. Fig.~\ref{fig:preprocessing} describes the evaluation results of these schemes. In this figure, the model w/-V has accuracy, precision, and recall at 93.849\%, 90.413\%, and 89.742\%, respectively. These results outperformed the model w/o-V, which has accuracy, precision, and recall at 72.163\%, 58.911\%, and 58.638\%, respectively. The DNN and DT schemes with Value features achieve accuracies of 85.87\% and 72.48\%, respectively. They are higher than that of the CNN without Value feature at 72.16\%, but lower than that of CNN with Value feature at 93.85\%. 
Especially, Fig.~\ref{fig:wop_wp} provides detailed information for all types of attacks and normal behavior. In Fig.~\ref{fig:wop_wp}, we can see that the model w/o-V cannot detect DoS and FoT attacks because it classifies all samples of DoS and FoT attacks into normal behavior. In contrast, the model w/-V can detect these types of attacks with high accuracy at about 97\% for DoS detection and 100\% for FoT detection. This is because the value feature is essential to support the learning models to detect many types of important attacks.

\subsubsection{Accuracy Analysis}
\begin{figure*}[!t]
    \begin{subfigure}{0.5\linewidth}
        \centering
        \includegraphics[width=.8\linewidth]{Figs/wp_11.pdf}
        \caption{}
        \label{fig:wp}
    \end{subfigure}
    \begin{subfigure}{0.5\linewidth}
        \centering
        \includegraphics[width=.8\linewidth]{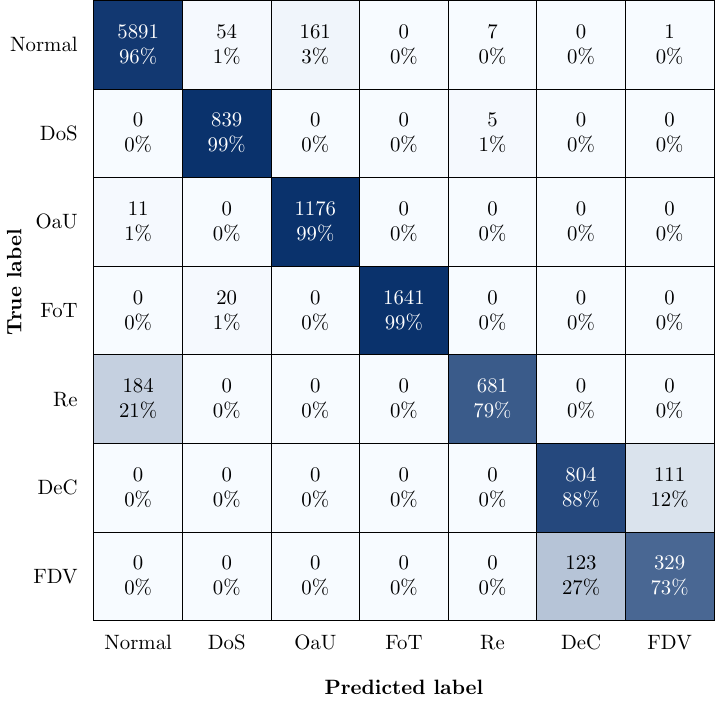}
        \caption{}
        \label{fig:CNN-5}
    \end{subfigure}
    \caption{The detection results of Centralized-CNN and Co-CNN-5 models. (a) Centralized-CNN w/-V. (b) Co-CNN-5 w/-V.}
    \label{fig:wp_CNN}
    \vspace*{-0.5cm}
\end{figure*}
\begin{figure*}
    \begin{subfigure}{0.5\linewidth}
        \centering
        \includegraphics[width=.8\linewidth]{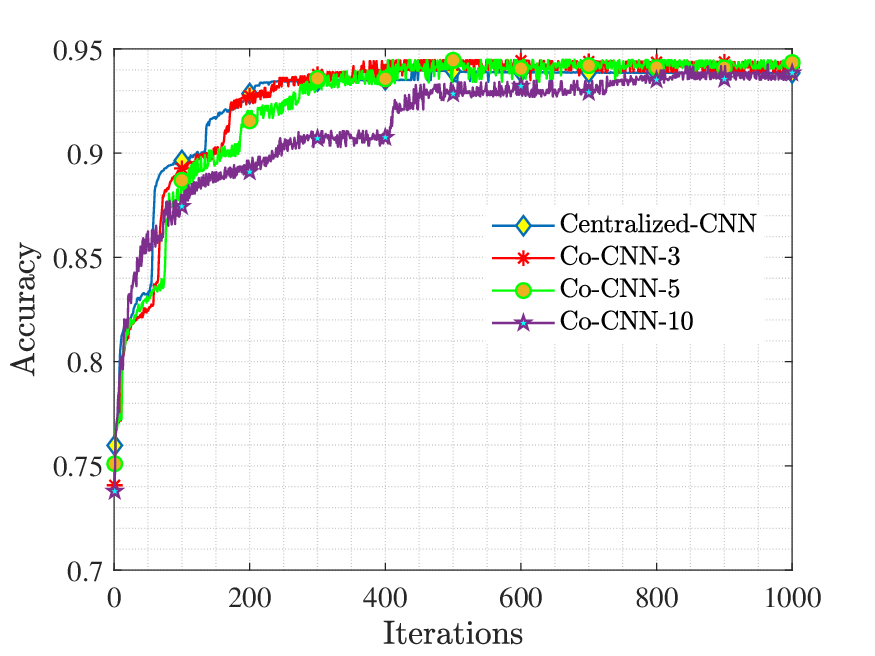}
        \caption{}
        \label{fig:acc}
    \end{subfigure}
    \begin{subfigure}{0.5\linewidth}
        \centering
        \includegraphics[width=.8\linewidth]{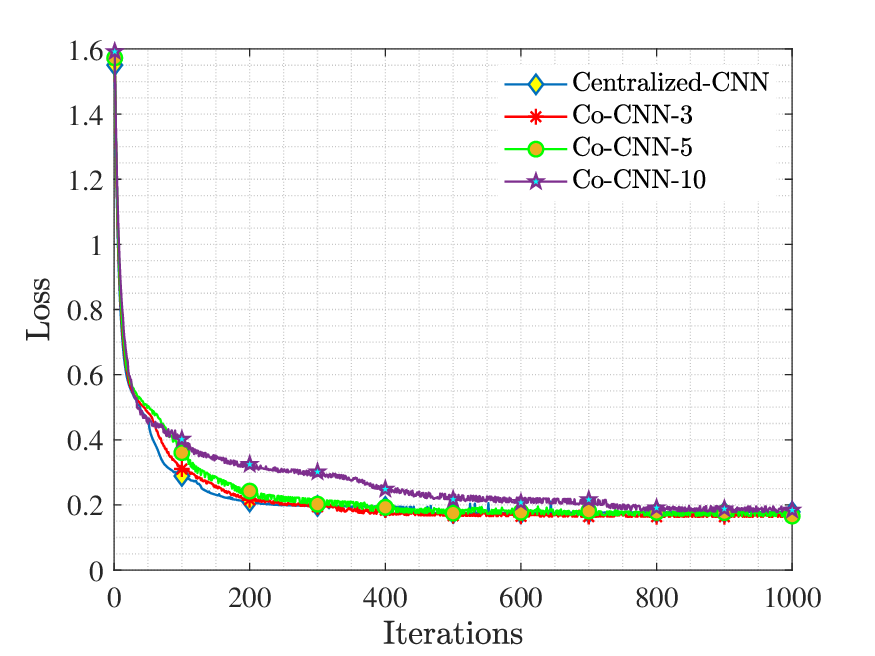}
        \caption{}
        \label{fig:loss}
    \end{subfigure}
    \caption{The convergence of accuracy and loss over iterations: (a) The accuracy over interactions, and (b) The loss over iterations.}
    \label{fig:acc_loss}
    \vspace*{-0.5cm}
\end{figure*}

In this section, we perform experiments to compare the performance results of the centralized model with our proposed model. The centralized model (Centralized-CNN) that we design can learn knowledge from all MNs for training and testing processes. Besides, we use different schemes of the collaborative learning model with 3 mining nodes (Co-CNN-3), 5 mining nodes (Co-CNN-5), and 10 mining nodes (Co-CNN-10). In each scheme, the collected datasets are divided equally among all mining nodes. To implement experiments, we first perform cyberattacks on transactions and SCs in our deployed private Ethereum platform to collect datasets from all MNs. In our proposed collaborative learning model, each MN uses its local dataset for both training and testing processes. However, in the training process, the MNs can exchange their trained models with others to improve their learning knowledge as well as the accuracy of attack detection. On the other hand, in the Centralized-CNN, all the local datasets of MNs will be gathered into a big dataset for its training and testing process. 

The performance results of two scenarios of preprocessing processes (i.e., without value feature (w/o-V) and with value feature (w/-V) with all schemes are also provided in Table~\ref{tab:simu_result_wop} and Table~\ref{tab:simu_result}. Table~\ref{tab:simu_result_wop} presents the performance of the simulation results of all schemes with the w/o-V preprocessing process. In Table~\ref{tab:simu_result_wop}, the accuracy, precision, and recall are nearly the same at around 72-73\%, 58-59\%, and 58-59\%, respectively. In contrast, in Table~\ref{tab:simu_result}, we can observe that the performance of all schemes with the w/-V preprocessing process outperforms those w/o-V preprocessing process at about 93-94\%, 90-91\%, and 89-90\% in accuracy, precision, and recall, respectively. In detail,  we first can see in Table~\ref{tab:simu_result} 
that the performance results of our proposed models are nearly the same as the Centralized-CNN. However, in some MNs, such as MN-5 of the Co-CNN-5, the accuracy, precision, and recall are higher than those of the Centralized-CNN at around 0.6\%, 0.6\%, and 0.7\%, respectively. Specifically, Fig.~\ref{fig:wp_CNN} provides detailed information for each type of attack of the Centralized-CNN and MN-5 of Co-CNN-5. These figures show that the misdetection of MN-5 of the Co-CNN-5 is dramatically reduced compared to the Centralized-CNN. In detail, the misdetection of the MN-5 from Normal to DoS is at 0.88\%, which is smaller than that of the Centralized-CNN at 1.14\%. Similarly, the misdetection of the MN-5 from OaU to Normal is at 0.926\% of total samples of OAU, which is smaller than that of the Centralized-CNN at 3.89\%.

\subsubsection{Convergence Analysis}
\begin{table*}
\centering
\caption{Real-time experiment results.}
\label{tab:real_result}
\begin{subtable}{\textwidth}
\caption{Centralized-CNN and Co-CNN w/-V}
\label{tab:real_result_wp}
\resizebox{\textwidth}{!}{
\begin{tabular}{|l|c|c|c|c|c|c|c|c|c|c|c|c|c|c|c|} 
\hline
\multirow{2}{*}{~ ~} & \multicolumn{5}{c|}{\textbf{Centralized-CNN}} & \multicolumn{5}{c|}{\textbf{Co-CNN-3}} & \multicolumn{5}{c|}{\textbf{Co-CNN-5}} \\ 
\cline{2-16}
 & \textbf{MN-1} & \textbf{MN-2} & \textbf{MN-3} & \textbf{MN-4} & \textbf{MN-5} & \textbf{MN-1} & \textbf{MN-2} & \textbf{MN-3} & \textbf{MN-4} & \textbf{MN-5} & \textbf{MN-1} & \textbf{MN-2} & \textbf{MN-3} & \textbf{MN-4} & \textbf{MN-5} \\ 
\hline
\textbf{Accuracy} & 89.603 & 89.542 & 89.668 & 89.702 & 89.291 & 88.663 & 88.582 & 88.655 & 88.794 & 88.471 & 90.928 & 90.896 & 90.957 & 91.061 & 90.614 \\ 
\hline
\textbf{Precision} & 76.851 & 75.806 & 76.956 & 76.690 & 75.582 & 76.755 & 75.872 & 76.845 & 77.191 & 75.912 & 80.192 & 78.835 & 80.469 & 80.846 & 78.576 \\ 
\hline
\textbf{Recall} & 76.858 & 77.117 & 76.888 & 76.767 & 76.822 & 78.523 & 78.939 & 78.531 & 78.563 & 78.724 & 78.870 & 79.044 & 78.757 & 78.762 & 78.747 \\
\hline
\end{tabular}
}
\end{subtable}
\hfill

\begin{subtable}{\textwidth}
\caption{Centralized-CNN and Co-CNN w/o-V}
\label{tab:real_result_wop}
\resizebox{\textwidth}{!}{
\begin{tabular}{|l|c|c|c|c|c|c|c|c|c|c|c|c|c|c|c|} 
\hline
\multirow{2}{*}{~ ~} & \multicolumn{5}{c|}{\textbf{Centralized-CNN}} & \multicolumn{5}{c|}{\textbf{Co-CNN-3}} & \multicolumn{5}{c|}{\textbf{Co-CNN-5}} \\ 
\cline{2-16}
 & \textbf{MN-1} & \textbf{MN-2} & \textbf{MN-3} & \textbf{MN-4} & \textbf{MN-5} & \textbf{MN-1} & \textbf{MN-2} & \textbf{MN-3} & \textbf{MN-4} & \textbf{MN-5} & \textbf{MN-1} & \textbf{MN-2} & \textbf{MN-3} & \textbf{MN-4} & \textbf{MN-5} \\ 
\hline
\textbf{Accuracy} & 65.877 & 65.780 & 65.643 & 66.312 & 65.734 & 66.804 & 66.640 & 66.569 & 67.115 & 66.797 & 65.606 & 65.579 & 65.512 & 66.212 & 65.830 \\ 
\hline
\textbf{Precision} & 47.263 & 46.105 & 46.739 & 47.544 & 46.012 & 51.442 & 50.024 & 51.116 & 51.641 & 50.433 & 44.668 & 44.078 & 44.534 & 44.994 & 44.141 \\ 
\hline
\textbf{Recall} & 51.383 & 51.576 & 51.434 & 51.318 & 51.427 & 49.670 & 49.888 & 49.586 & 49.557 & 49.625 & 48.447 & 48.544 & 48.381 & 48.372 & 48.518 \\
\hline
\end{tabular}
}
\end{subtable}
\end{table*}
\begin{figure*}
    \centering
    \begin{subfigure}{0.24\linewidth}
        \includegraphics[width=\linewidth]{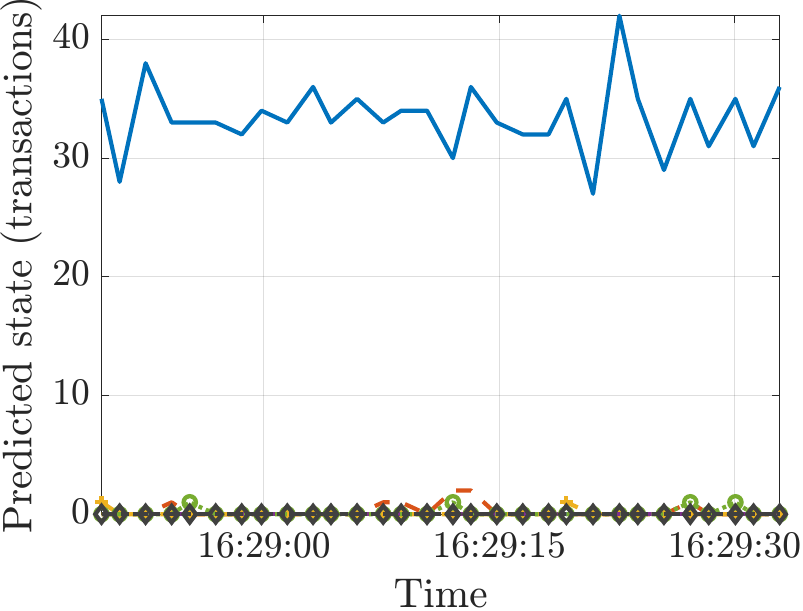}
        \caption{Normal state}
        \label{fig:realdec_1}
    \end{subfigure}
    \hfill
    \begin{subfigure}{0.24\linewidth}
        \includegraphics[width=\linewidth]{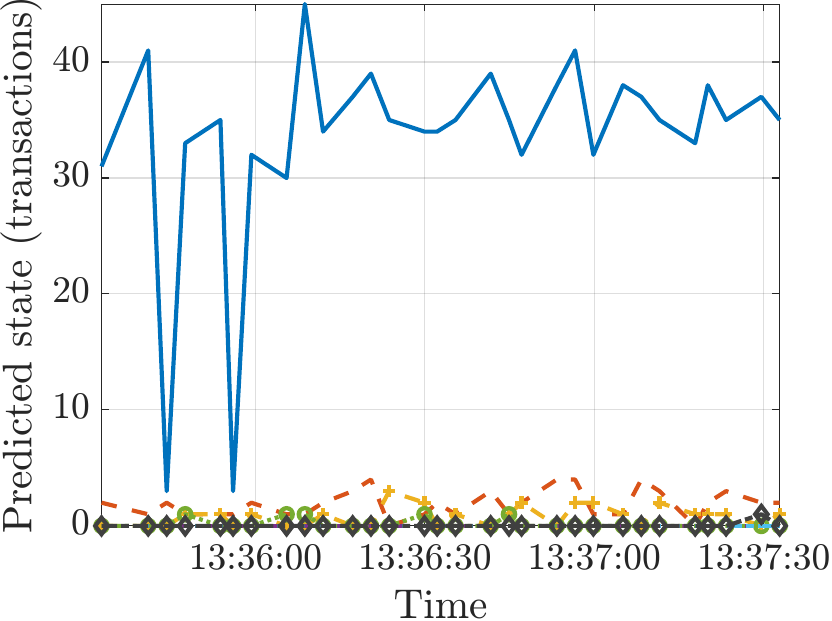}
        \caption{Re attack state}
        \label{fig:realdec_2}
    \end{subfigure}
    \hfill
    \begin{subfigure}{0.24\linewidth}
        \includegraphics[width=\linewidth]{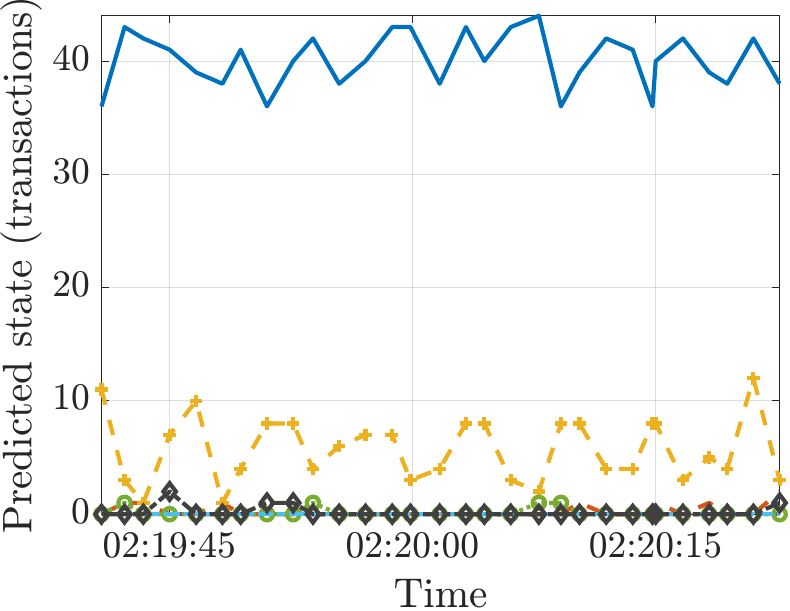}
        \caption{OaU attack state}
        \label{fig:realdec_3}
    \end{subfigure}
    \hfill
    \begin{subfigure}{0.24\linewidth}
        \includegraphics[width=\linewidth]{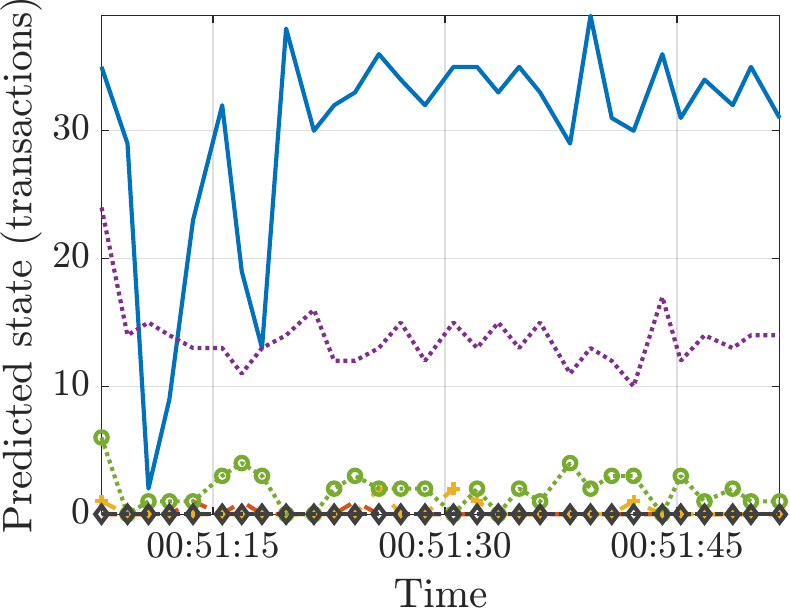}
        \caption{FoT attack state}
        \label{fig:realdec_4}
    \end{subfigure}
    \hfill
    \begin{subfigure}{0.24\linewidth}
        \includegraphics[width=\linewidth]{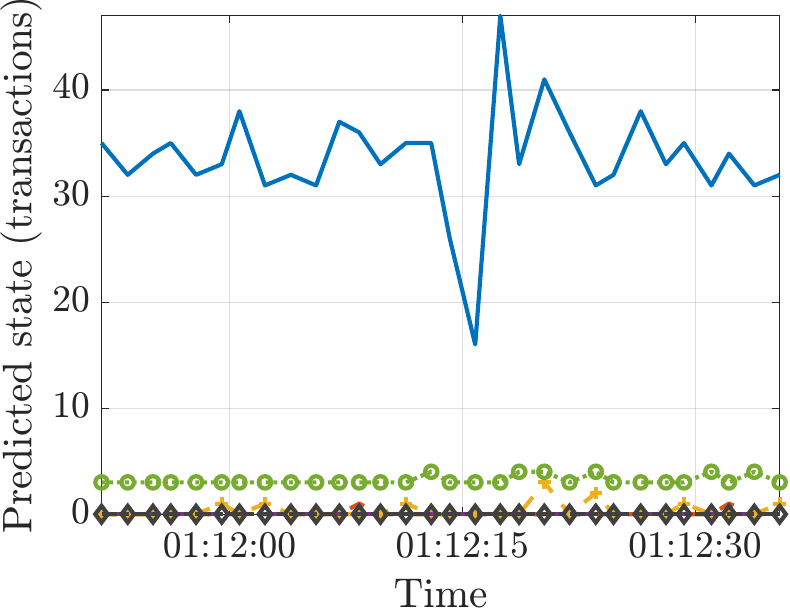}
        \caption{DoS attack state}
        \label{fig:realdec_5}
    \end{subfigure}
    \hfill
    \begin{subfigure}{0.24\linewidth}
        \includegraphics[width=\linewidth]{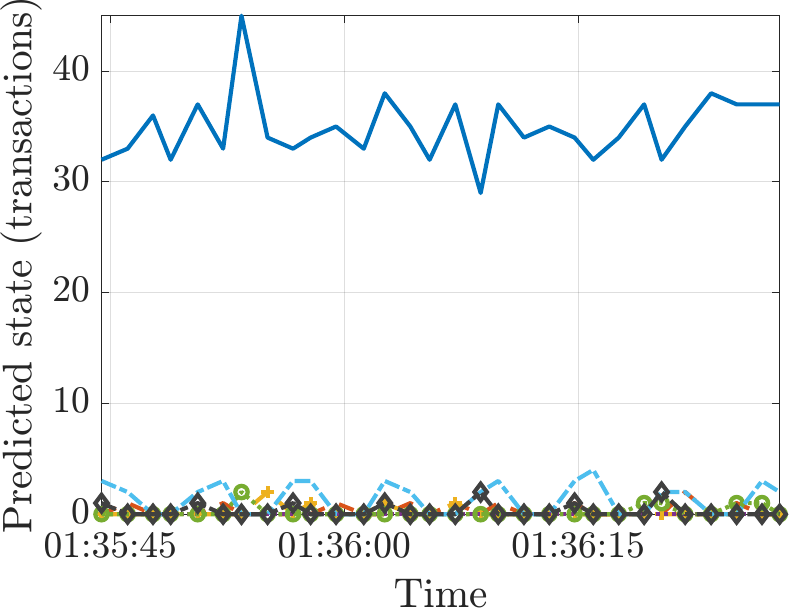}
        \caption{DeC attack state}
        \label{fig:realdec_6}
    \end{subfigure}
    \hfill
    \begin{subfigure}{0.24\linewidth}
        \includegraphics[width=\linewidth]{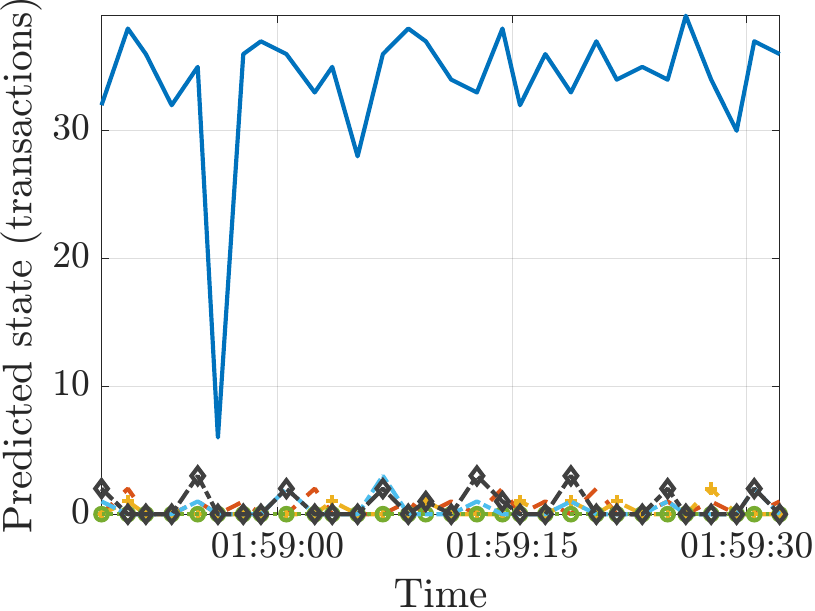}
        \caption{FDV attack state}
        \label{fig:realdec_7}
    \end{subfigure}
    \hfill
    \begin{subfigure}{\linewidth}
        \centering
        \includegraphics[width=0.5\linewidth]{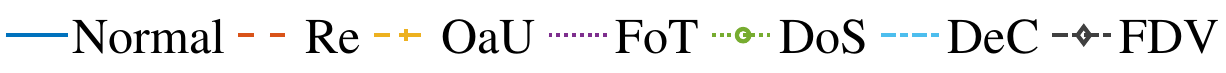}
    \end{subfigure}
    \vspace*{-0.3cm}
    \hfill
    \caption{Real-time cyberattack detection: proposed Co-CNN-5 model in Ethereum node 1.}
    \label{fig:realdec}
    \vspace*{-0.5cm}
\end{figure*}
In this section, we compare the convergence of different models, i.e., the Centralized-CNN, and the collaborative model with 3, 5, and 10 mining nodes. Fig.~\ref{fig:acc_loss} describes the accuracy and loss of these models in 1,000 iterations. In general, all of the models converged after about 800 iterations in terms of accuracy and loss. While the accuracies of Centralized-CNN, Co-CNN-3, and Co-CNN-5 models quickly reach the convergence after 400 iterations at about 93\%, the accuracies of Co-CNN-10 need about 800 iterations to converge and reach 93\%. The same trends happen with the loss. This is because the number of samples of each MN in Co-CNN-10 is much smaller than those of other models, while the number of workers is higher than those of other models. Thus, Co-CNN-10 needs more time to exchange learning knowledge with other models. It finally reaches convergence after about 800 iterations and has accuracies nearly the same as other models.

\subsubsection{Real-time Attack Detection}

In this section, we consider a practical scenario by evaluating the performance of the system in real-time cyberattack scenarios. To do this, we first take the trained models from all schemes (noted that the trained modes are trained in the schemes as in the accuracy analysis, i.e., Centralized-CNN, Co-CNN-3, Co-CNN-5). There are 5 blockchain nodes participating in these experiments and they join a private Ethereum network as described in the above section. After the learning models are trained, they are deployed on MNs. In the experiments, both two cases with value and without value preprocessing processes are considered. In real-time scenarios, both normal and attack samples continuously come to the blockchain node. Thus, the BCEC has to collect all the transaction traffic in 3~seconds into a package and then convert them into images. All processes including preprocessing (i.e., converting samples into images) and processing (i.e., model prediction) must be completed within 3~seconds before the next package comes.

Table~\ref{tab:real_result} presents the performance of Co-CNN-3, Co-CNN-5, and Centralized-CNN models in two cases of preprocessing. In general, we can observe in Table~\ref{tab:real_result_wp} that the performance of these models in accuracy, precision, and recall w/-V in the preprocessing process is at about 88-91\%, 76-80\%, and 77-79\%, respectively. These results outperform those of the w/o-V in the preprocessing process with accuracy, precision, and recall at about 65-66\%, 44-51\%, and 48-51\%, respectively. In addition, when we compare the same case w/-V in the preprocessing process of the simulation as in Table~\ref{tab:simu_result} and the real-time experimental results as in Table~\ref{tab:real_result_wp}, we can observe that the accuracy, precision, recall of the real-time experimental results are little smaller than those of simulation results about 3\%, 10\%, and 11\%, respectively. This is because, in simulation, we implement multiple types of attacks on the blockchain system and then collect data to have enough samples for the dataset to train the model. However, in real-time scenarios, some attack types, such as Re, DeC, and FDV, rarely appear during the experiment.
Thus, it makes it more difficult for the learning models to detect them in real-time. Specifically, we can observe in Table~\ref{tab:real_result_wp} that MN-4 of Co-CNN-5 has higher performance in accuracy, precision, and recall than MN-4 of the Centralized-CNN about 1.3\%, 4\%, and 2\%, respectively. Therefore, in real-time detection scenarios, our proposed model still demonstrates better performance in detecting attacks than in simulation.
\begin{figure}
    \centering
    \includegraphics[width=\linewidth]{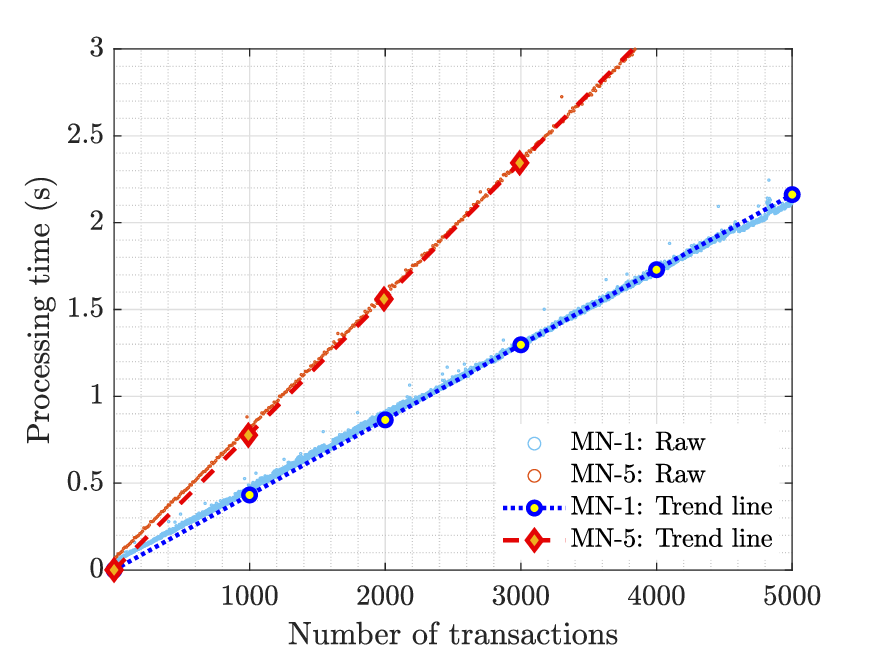}
    \caption{Throughput of proposed Co-CNN-5 model in two computer configurations.}
    \label{fig:throughput}
    \vspace*{-0.5cm}
\end{figure}
\subsubsection{Real-time Monitoring and Detection}

Fig.~\ref{fig:realdec} shows the real-time cyberattack monitoring from the output of our proposed model Co-CNN-5 in Ethereum node 1. In these figures, the normal and each type of attack are displayed in different lines. Fig.~\ref{fig:realdec_1} displays the normal state of the system with the high value of the predicted normal state over time. We can observe that in the normal state, the predicted states of all types of attacks are nearly 0. When a type of attack happens, the predicted state of that attack will increase, e.g., the FoT attack state as in Fig.~\ref{fig:realdec_4}. As described in the previous section, in real-time scenarios, Re, Dec, and FDV attack states have a small number of attack samples. Therefore, their predicted states in Fig.~\ref{fig:realdec_2}, Fig.~\ref{fig:realdec_6} and Fig.~\ref{fig:realdec_7} do not have high values. However, our proposed model can still detect all of the attacks in real-time with high accuracy at 91\%.

\subsubsection{Processing Time}

Fig.~\ref{fig:throughput} describes the processing time of two MNs with the same Co-CNN-5 model. We can observe in Fig.~\ref{fig:throughput} that when the number of transactions increases, the processing time of both MNs also linearly increases. However, there is a different capacity between the two MNs. In detail, while MN-5 can process about 1,100 transactions per second, the number of transactions that MN-1 can process is around 2,150 transactions per second. This is because of the different types of computer configuration between the two MNs described in section~\ref{sec:exper_setup}. However, in the mainnet of the Ethereum system, the maximum recorded number of transactions is 93.01 per second~\cite{Ether_transaction_daily}. Therefore, the capacity of our proposed system can be well-adapted to detect attacks on the mainnet Ethereum system.  

\section{Conclusion}
\label{sec:Conc}
In this work, we developed a collaborative learning model that can efficiently detect malicious attacks in transactions and SCs in a blockchain network. To do this, we implemented a private Ethereum network in our laboratory. We then performed attacks in transactions and SCs of that network for analysis. Next, we analyzed the transaction data and extracted the important features (i.e., Bytecode and value) to build the dataset. After that, we converted the dataset into grey images to train and evaluate the performance of our proposed model. In our proposed model, a learning node can detect the attacks in transactions and SCs of a blockchain network and receive and aggregate learning knowledge (i.e., trained models) from other learning nodes to improve the accuracy of detection. In this way, our proposed model does not expose the local data of learning nodes over the network, thereby protecting the privacy of the local data of learning nodes. Both simulation results and real-time experimental results showed the efficiency of our proposed model in detecting attacks. In the future, we will continue studying to develop other methods for detecting attacks in various kinds of networks.


\bibliographystyle{IEEEtran}
\bibliography{reference2}

\end{document}